\documentclass[authoryear, preprint, times]{elsarticle}
\usepackage{natbib}
\usepackage{graphicx}
\usepackage{amssymb}
\usepackage[version=3]{mhchem}  % formatting of chemical formulas

\usepackage{xcolor}

\usepackage{ulem}

\journal{Planetary and Space Sciences}

\begin{document}

\begin{frontmatter}

\title{Hot Oxygen and Carbon Escape from the Martian Atmosphere}

\author[iwf]{H. Gr\"oller}
\author[iwf]{H. Lichtenegger\corref{cor1}}
\ead{herbert.lichtenegger@oeaw.ac.at}
\author[iwf]{H. Lammer}
\author[inasan]{V. I. Shematovich}

\address[iwf]{Space Research Institute, Austrian Academy of Sciences, Schmiedlstrasse 6, A-8042 Graz, Austria}
\address[inasan]{Institute of Astronomy, Russian Academy of Sciences, 48 Pyatnitskaya St. 119017, Moscow, Russia}

\cortext[cor1]{Principal corresponding author}

\begin{abstract}
The escape of hot O and C atoms from the present martian atmosphere during low and high solar activity
conditions has been studied with a Monte-Carlo model. The model includes the initial
energy distribution of hot atoms, elastic, inelastic, and quenching collisions between the suprathermal  atoms
and the ambient cooler neutral atmosphere, and applies energy dependent total and differential cross sections for
the determination of the collision probability and the scattering angles.
The results yield a total loss rate of hot oxygen of $2.3-2.9\times 10^{25}\,{\rm s}^{-1}$ during low and high solar activity conditions and is mainly due to dissociative recombination of \ce{O2+} and \ce{CO2+}. The total loss rates of
carbon are found to be $0.8$ and $3.2\times 10^{24}\,{\rm s}^{-1}$ for low and high solar activity, respectively, with photodissociation of CO being the main source.
Depending on solar activity, the obtained carbon loss rates are up to $\sim 40$ times higher than the \ce{CO2+} ion loss rate inferred from Mars Express ASPERA-3 observations. Finally, collisional effects above the exobase reduce the escape rates by about 20-30\% with respect to a collionless exophere.
\end{abstract}


\end{frontmatter}

\section{Introduction}

The dynamics of the upper atmosphere and its interaction with the solar radiation and plasma
environment are important clues for the stability and long-term evolution of terrestrial atmospheres.
In a recent study \cite{lammer_outgassing_2013} showed that the evolution and
escape of the martian \ce{CO2} atmosphere and the planet's \ce{H2O} inventory can be
separated from the planet's origin into an early and a later evolutionary epoch.
The first epoch started after Mars' formation and lasted $\leq 500\,{\rm Myr}$ \citep{lammer_outgassing_2013}.
Because of the high soft X-ray and EUV flux of the young Sun
\citep[e.g.][]{ribas_evolution_2005, claire_evolution_2012} and Mars' low gravity the
nebula-based and outgassed hydrogen-rich protoatmosphere was lost by
hydrodynamic blow-off of H atoms and strong thermal escape rates of
dragged heavier species such as O and C atoms within several $10^4$ years
\citep[this issue]{erkaev_PSS_2013}. These recent results are in support of
the hypotheses of \cite{tian_thermal_2009} that early Mars could not build up a dense
\ce{CO2} atmosphere during the early Noachian.

However, after the planet lost its protoatmosphere
the atmospheric escape rates were most likely balanced with the volatiles (\ce{H2O}, \ce{CO2}, etc.)
which have been outgassed by volcanic activity and delivered by impacts until the activity
of the young Sun decreased, so that the atmospheric sources could dominate over
the losses $\sim 4 - 4.2\,{\rm Gyr}$ ago \citep{lammer_outgassing_2013}. If a secondary outgassed \ce{CO2} atmosphere of several
$100\,{\rm mbar}$ was indeed present at the end of the Noachian, its evolution during the second epoch which
lasted from the end of the Noachian until today should have been
determined by a complex interplay of various nonthermal atmospheric escape processes,
impacts, carbonate precipitation, and serpentinization during the Hesperian and Amazonian epochs which led
to the present day surface pressure \citep{chassefiere_serpentinization_2011,chassefiere_redox_2011}.

Atmospheres may get eroded through the loss
of light particles like H or He, when their thermal velocities exceed the escape velocity at the exobase.
Heavier species like O or C, are stronger bound to the planet and can therefore either directly escape through nonthermal
processes like dissociative recombination of molecular ions or may form extended coronae which in turn can be
eroded by the solar wind. From Mars Express ASPERA-3 data \cite{barabash_martian_2007} estimated
the present escape of \ce{CO2+} molecular ions and yield low escape rates
of $\sim 8 \times 10^{22}\,\rm s^{-1}$. Based on this rate these authors estimated that the fraction
of \ce{CO2+} molecular ions lost to space since the end of the Noachian when the martian
dynamo stopped to work would represent an equivalent surface pressure of only about $\sim 0.2 - 4\,{\rm mbar}$.

\cite{lammer_outgassing_2013} compared these ion escape rates with atmospheric sputtering and exothermic
escape processes of heavy neutrals and suggested that the loss of photochemically produced
suprathermal atoms such as O, C may be responsible for the main loss of the martian \ce{CO2} atmosphere at present
and most likely also during the planet's history until thermal escape
took over.

The simulation of this nonthermal photochemically related escape of heavy atmospheric atoms at terrestrial planets is not only challenged by the physical and chemical complexity of the upper atmosphere but is additionally hindered by our sparse knowledge
of many of the processes which govern the direct escape of atmospheric constituents. Although Mars has been visited by many spacecraft for more than 40 years, observational data of the hot particle corona at Mars are still rather limited \citep[e.g.][]{leblanc_spicam_2006, lichtenegger_temperature_2007};
the only recent observation of the extended oxygen corona of Mars was made with the Alice far-ultraviolet
imaging spectrograph during the Rosetta swing-by of Mars \citep{feldman_rosetta_2011}.
Therefore, numerical modeling of exosphere densities and of escape rates are still little constrained and suffer from the various
assumptions made by different authors.

The aim of this work is to obtain a better understanding on the escape of hot carbon and oxygen atoms from present Mars.
For this purpose we consider a number of possible sources of hot O and C atoms in the thermosphere, calculate the production
rate profiles and the energy distribution functions at the exobase for low and high solar activity conditions and discuss the
associated escape fluxes and exosphere density profiles. Because there are little observational hints of the rotational and
vibrational excitation distributions of the particles and because these effects have negligible influence on the escape rates
\citep{groller_venus_2010}, we assume all particles in the rotational and vibrational ground states.

The simulations are based -- if available --  on the most recent data for both the atmospheric input and the collision cross sections. The model is limited by various circumstances: (a) the background atmosphere is represented by 1D ion and neutral density profiles (i.e. all variables depend only on radial distance) assumed to be valid for 60$^{\circ}$ solar zenith angle, (b) the hot particles are treated as test particles, i.e. their influence on the background gas is neglected, (c) due to the lack of data, some types of collisions are simulated by using approximate values of the parameters (see Tables 9 and 10).

\section{Hot Atom Modeling}
%--------------------------
\subsection{Monte Carlo model}
%-----------------------------
The simulation of the hot particle corona is initiated by calculating for a specific reaction the corresponding velocity
distribution of the products at discrete altitudes between 130 km and 450 km above the surface. These products are represented
by a number of particles (typically $\sim 10^4$ per 2.5 km altitude step) and their 3-dimensional motion is followed through the thermosphere up to 450 km in the gravitational field of Mars. On their way the hot particles can interact with the background neutral atmosphere via elastic, inelastic and quenching collisions and will loose on average part of their initial energy. Background particles which acquire an energy exceeding the thermal velocity of the atmospheric species upon collisions are also treated as new hot particles. The collision probability and energy transfer is calculated by means of total and differential cross sections. At the exobase (for comparison with other models) and at 380 km altitude,
the energy distribution
function of the suprathermal particles is determined which in turn serves as input for the exosphere density calculations.
A more detailed description of the Monte-Carlo model can be found in \cite{groller_venus_2010} and \cite{groller_hot_2012}.

Although some 3D simulation results of the martian thermosphere have been published in the recent past \citep{valeille_study_2010,yagi_icarus_2012}, our background thermosphere model is assumed to be spherically symmetric for various reasons:
(a) we do not focus on local variations such as solar zenith angle effects;
(b) the primary aim of our paper is the study of a number of chemical reactions producing hot particles in view of their relevance for escape
as well as the treatment of the various types of collisions on the basis of total and differential cross sections rather than on hard sphere
or universal potential models, and this analysis can be done most easily in 1D;
(c) to facilitate the comparison with older models which are to a large extend 1D models;
(d) due to the lack of sufficient observations, realistic 3D thermosphere models of Mars are yet difficult to construct.

\subsection{Input Profiles}
%--------------------------
The neutral and ion density as well as the temperatures for both high and low solar activity are taken from \cite{fox_photochemical_2009} for a solar zenith angle of $60\,^\circ$. In their model, the ambipolar diffusion for ions and the molecular and eddy diffusion for neutrals are included as well as more than 220 reactions. The altitude dependent density profiles of the background atmosphere are based on measurements of the Viking 1 neutral mass spectrometer together with simulations of the Mars Thermospheric General Circulation Model (MTGCM).
According to \cite{fox_photochemical_2009}, the O$_2^+$ density profiles measured by Viking can only be reproduced by imposing an upward velocity boundary condition ("eroded" ionospheres); therefore, only these eroded models are used in the present paper as input for the ion densities.

The solar flux is taken from the \textit{Solar Irradiance Platform (SIP)} version 2.36 \citep{tobiska_solar2000_2004}, where we have chosen DOY 32 of 2002 to represent high solar activity (HSA) and DOY 110 of 1997 to represent low solar activity (LSA). For these dates, HSA and LSA correspond to a solar flux of $F_{10.7} = 245.6$ and $F_{10.7} = 69.7$, respectively.
Finally, for each considered neutral element in the atmosphere, the corresponding photo cross sections are taken from the \textit{photo ionization/dissociation rates} database of \cite{huebner_solar_1992}.

% ****************************************
% ****     Sources of Hot O Atoms     ****
% ****************************************
\subsection{Sources of Hot O Atoms}
%----------------------------------
Although dissociative recombination (DR) of \ce{O2+} ions is supposed to be the main source of hot oxygen atoms \citep[e.g.][]{mcelroy_science_1972, nagy_hot_1988, kim_solar_1998, fox_photochemical_2009, groller_venus_2010, groller_hot_2012}, we have also considered some additional possible sources of hot O provided that the corresponding density profiles are available. These additional reactions are listed in Tables 2 to 6.

\subsubsection{Dissociative and Radiative Recombination}
%-------------------------------------------------------
Dissociative recombination of ground-state \ce{O2+} proceeds via the five exothermic channels shown in Table \ref{tab:DR_O2plus_branches}.\
\begin{table}[h]
    \centering
    \setlength{\belowcaptionskip}{7pt}
    \caption{Excess energies \citep[e.g.][]{fox_photochemical_2009} and branching ratios \citep{Peverall_et_al_2001} for dissociative recombination of \ce{O2+(X^2\Pi_g)}. The branching ratios correspond to approximately zero collision energy.}
    \label{tab:DR_O2plus_branches}
    \begin{tabular}{ llcc } %\toprule
    \hline
   \multicolumn{1}{l}{\bf Reaction}   &   \multicolumn{1}{l}{\bf Channels} &   \multicolumn{1}{c}{\bf Excess Energy [eV]}  &   \multicolumn{1}{c}{\bf Branching Ratios [\%]} \\
    \hline
    \multicolumn{1}{l}{O$_2^+$ + e}  &   \multicolumn{1}{l}{O($^3$P) + O($^3$P)}  & 6.99 &  20.4 \\
                                     &   \multicolumn{1}{l}{O($^3$P) + O($^1$D)}  & 5.02 &  44.0 \\
                                     &   \multicolumn{1}{l}{O($^1$D) + O($^1$D)}  & 3.05 &  31.5 \\
                                     &   \multicolumn{1}{l}{O($^3$P) + O($^1$S)}  & 2.80 &  0.0  \\
                                     &   \multicolumn{1}{l}{O($^1$D) + O($^1$S)}  & 0.83 &  4.1  \\
    \hline %\bottomrule
    \end{tabular}
\end{table}
%\noindent
\begin{table}[b]
    \centering
    \setlength{\belowcaptionskip}{7pt}
    \caption{Excess energies \citep[e.g.][]{fox_dissociative_2004} and branching ratios \citep{viggiano_rate_2005} for dissociative recombination of \ce{CO2+}; the uncertainties are given in parentheses.}
    \label{tab:DR_CO2plus_branches}
    \begin{tabular}{ llcc } %\toprule
    \hline
    \multicolumn{1}{l}{\bf Reaction}   &   \multicolumn{1}{l}{\bf Channels} &   \multicolumn{1}{c}{\bf Excess Energy [eV]}  &   \multicolumn{1}{c}{\bf Branching Ratios [\%]} \\
    \hline
    \multicolumn{1}{l}{\ce{CO2+ + e}}  &   \multicolumn{1}{l}{\ce{CO2(^1\Sigma_g)}}              & 13.78 &  0.0 (+ 2.0)   \\
                                       &   \multicolumn{1}{l}{\ce{C(^3P)} + \ce{O2(^3\Sigma_g)}} &  2.29 &  0.0 (+ 4.0)   \\
                                       &   \multicolumn{1}{l}{\ce{CO(^1\Sigma)} + \ce{O(^3P)}}   &  8.27 &  100.0 (- 6.0)  \\
                                       &   \multicolumn{1}{l}{\ce{C(^3P)} + 2\ce{O(^3P)}}        & -2.87 &  0.0   \\
    \hline %\bottomrule
    \end{tabular}
\end{table}
\noindent
The excited states O($^1$D) and O($^1$S) are $1.97\,{\rm eV}$ and $4.19\,{\rm eV}$ above the ground state O($^3$P), respectively.
For the ground vibrational state of the oxygen molecule \ce{O_2+($v=0$)}, the branching ratios are taken from \cite{Peverall_et_al_2001} for an electron-ion collision energy of nearly $0\,{\rm eV}$. There are more recent values published by \cite{Petrignani_et_al_2005_electron} but they state that ``The branching fractions and quantum yields do not entirely agree with earlier measurements from our group that were taken at energies between 0 and $40\,{\rm meV}$ in a cold ion beam, nor with measurements performed at ASTRID by
\cite{Kella_et_al_1997}.'' Therefore and because the listed branching ratios are comparable to those of \cite{Kella_et_al_1997}, the values from \cite{Peverall_et_al_2001} are chosen.

\begin{table}[h!]
    \centering
    \setlength{\belowcaptionskip}{7pt}
    \caption{Excess energies and branching ratios \citep{Vejby-Christensen_et_al_1998} for dissociative recombination of \ce{NO+(X^1\Sigma+)}; the latter are valid for zero electron-ion collision energy.}
    \label{tab:DR_NOplus_branches}
    \begin{tabular}{ llcc } %\toprule
    \hline
    \multicolumn{1}{l}{\bf Reaction}   &   \multicolumn{1}{l}{\bf Channels} &   \multicolumn{1}{c}{\bf Excess Energy [eV]}  &   \multicolumn{1}{c}{\bf Branching Ratios [\%]} \\
    \hline
    \multicolumn{1}{l}{NO$^+$ + e}  &   \multicolumn{1}{l}{N($^4$S) + O($^3$P)}  & 2.78 &  15.0   \\
                                  &   \multicolumn{1}{l}{N($^4$S) + O($^1$D)}  & 0.81 &   0.0   \\
                                  &   \multicolumn{1}{l}{N($^2$D) + O($^3$P)}  & 0.39 &  85.0   \\
    \hline %\bottomrule
    \end{tabular}
\end{table}
The dissociative recombination of \ce{CO2+} and the corresponding excess energies and branching ratios are listed in Table \ref{tab:DR_CO2plus_branches}.
The branching ratios are taken from \cite{viggiano_rate_2005}, where the uncertainties are given in parentheses.
Of all channels, only the third and fourth produce atomic oxygen, although the latter is of no interest since it is an endothermic process
and its probability is close to zero.
\begin{table}[b!]
    \centering
    \setlength{\belowcaptionskip}{7pt}
    \caption{Excess energies \citep[e.g.][]{fox_photochemical_2001} and branching ratios \citep{rosen_absolute_1998} for dissociative recombination of \ce{CO+(X^2\Sigma+)} for collision energies of 0.0, 0.4, 1.0, and 1.5 eV.}
    \label{tab:DR_COplus_branches}
    \begin{tabular}{llccccc} %\toprule
    \hline
    \multicolumn{1}{l}{\bf Reaction}   &   \multicolumn{1}{l}{\bf Channels} &   \multicolumn{1}{c}{\bf Excess Energy [eV]}  &   \multicolumn{4}{c}{\bf Branching Ratios [\%]} \\
                                    &  &  & \multicolumn{1}{c}{0.0 eV}  &  \multicolumn{1}{c}{0.4 eV}  &  \multicolumn{1}{c}{1.0 eV}  &  \multicolumn{1}{c}{1.5 eV} \\
    \hline
    \multicolumn{1}{l}{CO$^+$ + e}  &  \multicolumn{1}{l}{C($^3$P) + O($^3$P)}  &  2.90 & 76.1  &  53.0  &  39.0  &  38.0  \\
                                  &  \multicolumn{1}{l}{C($^1$D) + O($^3$P)}  &  1.64 & 14.5  &  34.0  &  35.0  &  35.0  \\
                                  &  \multicolumn{1}{l}{C($^3$P) + O($^1$D)}  &  0.93 & 9.4  &  8.0  &  15.0  &  11.0  \\
                                  &  \multicolumn{1}{l}{C($^1$S) + O($^3$P)}  &  0.22 & 0.0  &  0.0  &  5.0  &  5.0  \\
                                  &  \multicolumn{1}{l}{C($^1$D) + O($^1$D)}  & -0.32 &   &  5.0  &  6.0  &  11.0  \\
                                  &  \multicolumn{1}{l}{C($^3$P) + O($^1$S)}  & -1.28 &  &    &    &  0.0  \\
    \hline %\bottomrule
    \end{tabular}
\end{table}
\noindent

\begin{table}[b!]
    \centering
    \caption{Possible sources of hot O atoms due to dissociative and radiative recombination and references for the rate coefficients.}\vspace{2mm}
    \begin{tabular}{@{} p{4.5cm}  p{6.5cm} @{}}
    \hline
    {\bf Source}     &   {\bf Rate coefficient [$\mathbf{cm^3\,s^{-1}}$]}   \\
    \hline
	\multicolumn{2}{c}{\bf Dissociative recombination}   \\
	\ce{O2+ + e -> O + O}   &   \cite{sheehan_dissociative_2004}  \\
	                         &   $\quad T \leq 1200\,{\rm K}: \alpha=2.40\times10^{-7}$, $\beta=-0.70$  \\
	                         &   $\quad T > 1200\,{\rm K}: \alpha=1.93\times10^{-7}$, $\beta=-0.61$  \\
	\ce{CO2+ + e -> CO + O}   &   \cite{viggiano_rate_2005}     \\
	                         &   $\quad \alpha=4.20\times10^{-7}$, $\beta=-0.75$  \\ %T > 1200\,{\rm K}:
	\ce{NO+ + e -> N + O}   &   \cite{sheehan_dissociative_2004}  \\
	                         &   $\quad T \leq 1200\,{\rm K}: \alpha=3.50\times10^{-7}$, $\beta=-0.69$  \\
	                         &   $\quad T > 1200\,{\rm K}: \alpha=3.02\times10^{-7}$, $\beta=-0.56$  \\
	\ce{CO+ + e -> C + O}   &   \cite{rosen_absolute_1998}     \\
	                         &   $\quad \alpha=2.75\times10^{-7}$, $\beta=-0.55$  \\ 
	\multicolumn{2}{c}{\bf Radiative recombination}   \\
	\ce{O+ + e -> O + h$\nu$}   &   \cite{nahar_electronion_1999}  \\
	                         &   $\quad \alpha=3.24\times10^{-12}$, $\beta=-0.66$  \\
    \hline
    \end{tabular}
    \label{tab:O_sources_recombination}
\end{table}
The channels for dissociative recombination of \ce{NO+} are given in Table \ref{tab:DR_NOplus_branches}. The branching ratios taken from \cite{Vejby-Christensen_et_al_1998} are valid for the vibrational ground state and for an electron-ion collision energy close to 0 eV.
The excited states O($^1$D) and N($^2$D) have energies of $1.97\,{\rm eV}$ and $2.38\,{\rm eV}$ above the corresponding ground states, respectively.

The excess energies and branching ratios for the dissociative recombination of \ce{CO+} are listed in Table \ref{tab:DR_COplus_branches}. The branching ratios for collision energies of 0.0, 0.4, 1.0, and $1.5\,{\rm eV}$ are taken from \cite{rosen_absolute_1998}.

The altitude dependent production rate $P(r)$ of the dissociative recombination of an ion can be calculated by
\begin{equation}\label{eq:DR_prod_rate}
  P(r) = n_{\rm i}(r) \ n_{\rm e}(r) \ k(T_{\rm e}(r)) ,
\end{equation}
where $n_{\rm i}(r)$ and $n_{\rm e}(r)$ are the altitude dependent ion and electron densities, respectively, and $k(T_{\rm e}(r))$ is the rate coefficient which depends on the electron temperature $T_{\rm e}(r)$. The value of $k$ is determined according to the Arrhenius equation as
\begin{equation}
  k(T_{\rm e}(r)) = \alpha \left( \frac{T_{\rm e}(r)}{300} \right)^\beta \qquad {\rm cm}^{3}\,{\rm s}^{-1}.
\end{equation}
The coefficients $\alpha$ and $\beta$ used for the dissociative recombination processes considered in this paper are listed in Table \ref{tab:O_sources_recombination}.

\begin{figure}[t]
	\begin{center}
	\includegraphics[width=0.65\columnwidth]{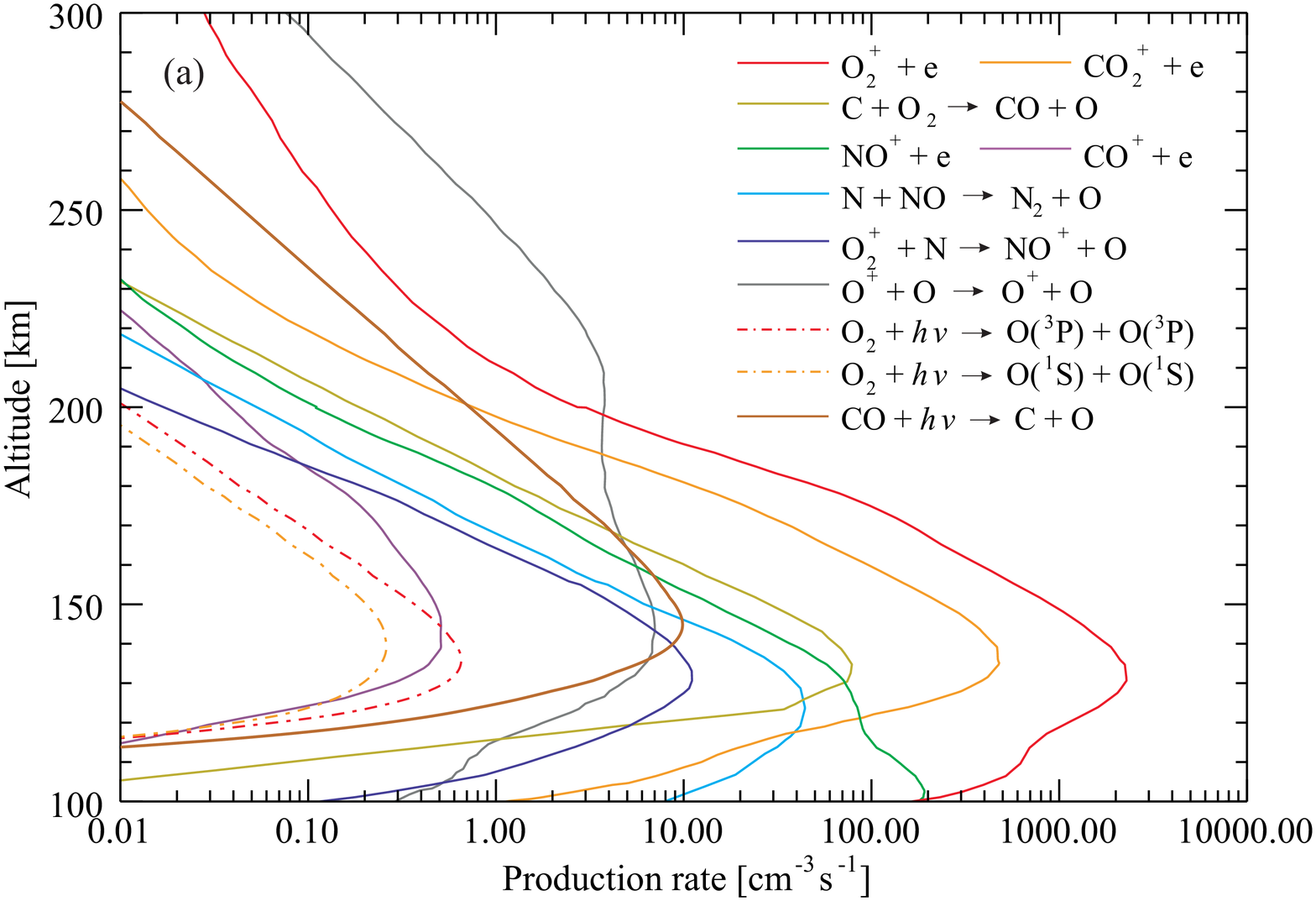}\vspace{5mm}
	\includegraphics[width=0.65\columnwidth]{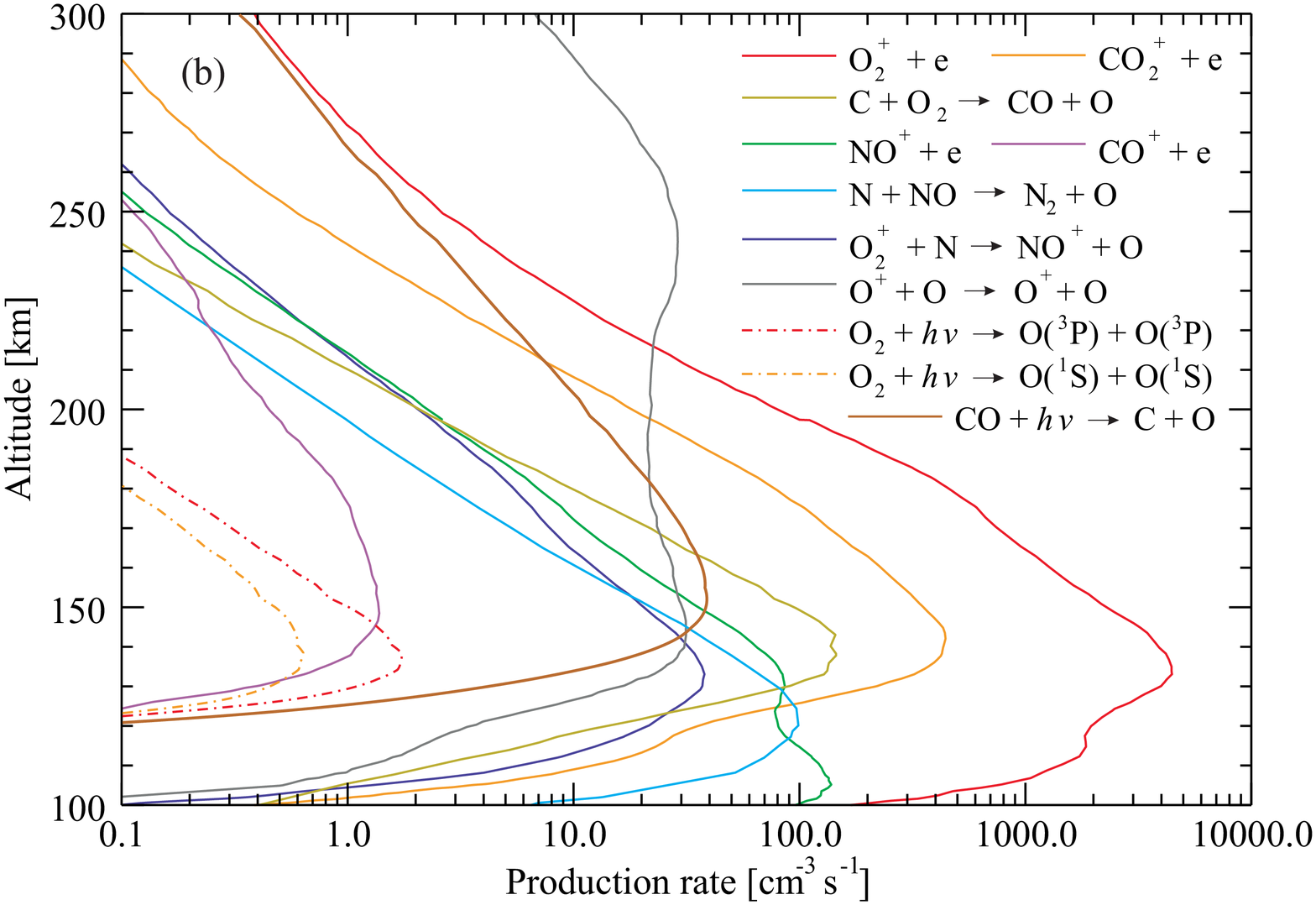}
	\caption{Production rates of hot O atoms for low (a) and high (b) solar activity due to various sources of oxygen.}
	\label{fig:production_rates_O}
	\end{center}
\end{figure}

Another possible source of hot O is radiative recombination of \ce{O+}
\begin{align*}
	\ce{O+(^4S) + e -> O(^3P) + $h\nu$},
\end{align*}
where the parameters $\alpha$ and $\beta$ for the rate coefficient (2) are taken from the UDFA06 database \citep{woodall_umist_2007} based on \cite{nahar_electronion_1999} and given in Table \ref{tab:O_sources_recombination}. However, radiative recombination appears to be rather inefficient, yielding maximum production rates of about $1\times 10^{-4}$ and $2\times 10^{-5}$ cm$^{-3}\,\rm{s}^{-1}$ for high and low solar activity, respectively.

The production rates as a function of altitude for the processes listed in Table \ref{tab:O_sources_recombination} are illustrated in Figure 1 for low (a) and high (b) solar activity.

\subsubsection{Photodissociation}
%--------------------------------
A further possible source of hot oxygen is photodissociation of \ce{O2}
\begin{align*}
	\ce{O2 + $h\nu$ -> O + O},
\end{align*}
for which an energy of at least $5.16\,{\rm eV}$ is needed (see \textit{Bond Dissociation Energies in Simple Molecules} - B. deB. Darwent. NSRDS-NBS 31,
(1970))\footnote{http://www.nist.gov/data/nsrds/NSRDS-NBS31.pdf}.

The altitude dependent production rate $P(r)$ for photodissociation of a molecular neutral $n$ can be obtained by the relation
\begin{equation}
  P(r) = n_n(r) \int\limits_{\lambda} \! \sigma_n^{\rm pd}(\lambda) F(r,\lambda) \, \mathrm{d}\lambda
\end{equation}
where $n_n(r)$ is the altitude dependent neutral density, $F(r,\lambda)$ the solar flux at the altitude $r$ and $\sigma_n^{\rm pd}(\lambda)$ the photodissociation cross section for the neutral at wavelength $\lambda$.
The photodissociation and absorption cross sections are taken from \cite{huebner_solar_1992} who provide them in the \textit{Photo Rate Coefficient Database}\footnote{\texttt{http://phidrates.space.swri.edu (retrieved Nov 2012)}}.

For low and hight solar activity, the production rate of photodissociation of \ce{O2} and \ce{CO} valid for a solar zenith angle
of 60$^\circ$ is illustrated in Figure 1. As can be seen, photodissociation of \ce{CO} -- though significant for the production of hot C (see Figure 2) -- is of little importance for the production of hot O.

\subsubsection{Chemical Reaction and Charge Transfer}
%----------------------------------------------------
The rate coefficient $k$ for a two-body chemical reaction as well as for charge transfer is determined by means of the Arrhenius equation
\begin{equation}
	k = \alpha \left( \frac{T}{300} \right)^\beta \exp \left( \frac{-\gamma}{T} \right) \qquad {\rm cm}^{3}\,{\rm s}^{-1},
\end{equation}
where $T$ is the gas temperature. The coefficients $\alpha$, $\beta$, and $\gamma$ are taken
from the UDFA06 database described in \cite{woodall_umist_2007} and are listed in Table \ref{tab:O_sources_reaction}.
The production rate for the charge transfer between an ionized and a neutral oxygen atom is calculated using a rate coefficient equal to $1.6 \times 10^{-11}\left(T_{\rm n} + T_{\rm i}\right)^{0.5}\,{\rm cm}^{3}\,{\rm s}^{-1}$ taken from \cite{terada_global_2002} with $T_{\rm n}$ and $T_{\rm i}$ being the neutral and ion temperature, respectively.
\begin{table}[h!]
    \small
    \begin{center}
    \setlength{\belowcaptionskip}{-7pt}
    \caption{Possible sources of hot O atoms with their corresponding coefficients $\alpha$, $\beta$, and $\gamma$ to calculate the rate coefficient $k$ in ${\rm cm}^{3}\,{\rm s}^{-1}$. The excess energies for chemical reactions are given for the ground state of the products, e.g. \ce{O($^3\rm{P}$)}, \ce{CO(X$^1\Sigma^+$)}.}
    \begin{tabular}{lccccc}
    \hline
    {\bf Source}     &   \multicolumn{3}{c}{\bf Rate coefficient [$\mathbf{cm^3\,s^{-1}}$]}   &  \multicolumn{2}{l}{\bf Production rate [$\mathbf{cm^3\,s^{-1}}$]} \\
         &  $\mathbf{\alpha}$   &  $\mathbf{\beta}$  &  $\mathbf{\gamma}$          &     {\bf HSA}    &     {\bf LSA}   \\
    \hline
	\multicolumn{4}{c}{\bf Chemical reaction}   \\
	\ce{C + NO -> CN + O + $1.22\,{\rm eV}$}   &   $6.00\times10^{-11}$ & $-0.16$ & $0.0$     &   $< 3 \times 10^{0}$   &   $< 2 \times 10^{0}$\\
	\ce{C + O2 -> CO + O(^1D) + $4.04\,{\rm eV}$}   &   $^a4.70\times10^{-11}$ & $-0.34$ & $0.0$   &  Figure 1b   &   Figure 1a  \\
	   &   $^b2.48\times10^{-12}$ & $1.54$ & $613.0$   &   Figure 1b   &   Figure 1a  \\	
	\ce{C + O2+ -> CO+ + O + $4.04\,{\rm eV}$}   &   $5.20\times10^{-11}$ & $0.00$ & $0.0$   &   $< 5 \times 10^{-1}$   &   $< 2 \times 10^{-1}$   \\
	\ce{C+ + O2 -> CO+ + O + $3.23\,{\rm eV}$}   &   $3.80\times10^{-10}$ & $0.00$ & $0.0$   &   $< 2 \times 10^{0}$   &   $< 10^{-1}$   \\
	\ce{H + CO2+ -> HCO+ + O + $0.83\,{\rm eV}$}   &   $2.90\times10^{-10}$ & $0.00$ & $0.0$   &   $< 3 \times 10^{-1}$   &   $< 2 \times 10^{0}$   \\
	\ce{He+ + CO -> C+ + O + He + $2.19\,{\rm eV}$}   &   $1.60\times10^{-9}$ & $0.00$ & $0.0$   &   $< 10^{-2}$   &   $< 10^{-2}$   \\
	\ce{He+ + CO2 -> CO+ + O + He + $1.04\,{\rm eV}$}   &   $8.70\times10^{-10}$ & $0.00$ & $0.0$   &     $< 10^{-2}$   &   $< 10^{-2}$   \\
	\ce{He+ + NO -> N+ + O + He + $3.51\,{\rm eV}$}   &   $1.40\times10^{-9}$ & $0.00$ & $0.0$   &     $< 10^{-2}$   &   $< 10^{-2}$   \\
	\ce{He+ + O2 -> O+ + O + He + $5.81\,{\rm eV}$}   &   $1.10\times10^{-9}$ & $0.00$ & $0.0$   &     $< 10^{-2}$   &   $< 10^{-2}$   \\
	\ce{N + NO -> N2 + O + $3.26\,{\rm eV}$}   &   $3.75\times10^{-11}$ & $0.00$ & $26.0$  &   Figure 1b   &   Figure 1a   \\
	\ce{N + O2+ -> NO+ + O + $4.17\,{\rm eV}$}   &   $1.80\times10^{-10}$ & $0.00$ & $0.0$  &   Figure 1b   &   Figure 1a   \\
	\ce{N+ + NO -> N2+ + O + $2.23\,{\rm eV}$}   &   $7.90\times10^{-11}$ & $0.00$ & $0.0$   &     $< 10^{-4}$   &   $< 3 \times 10^{-5}$   \\
	\ce{N+ + O2 -> NO+ + O + $6.65\,{\rm eV}$}   &   $2.63\times10^{-10}$ & $0.00$ & $0.0$   &     $< 10^{-2}$   &   $< 3 \times 10^{-3}$   \\
	\multicolumn{4}{c}{\bf Charge transfer}   \\
	\ce{O+ + H -> H+ + O + $0.02\,{\rm eV}$}   &   $5.66\times10^{-10}$ & $0.36$ & $-8.6$   &   $< 3 \times 10^{-2}$   &   $< 2 \times 10^{-1}$   \\
	\ce{O+ + O2 -> O2+ + O + $1.55\,{\rm eV}$}   &   $1.90\times10^{-11}$ & $0.00$ & $0.0$   &     $< 10^{0}$   &   $< 3 \times 10^{-2}$   \\
	\ce{O+ + O -> O+ + O + $0.00\,{\rm eV}$}   &    \multicolumn{3}{c}{$1.6 \times 10^{-11}\left(T_{\rm n} + T_{\rm i}\right)^{0.5}$}   &   Figure 1b  &
    Figure 1a   \\
    \hline
    \end{tabular}
    \label{tab:O_sources_reaction}
    \end{center}
    \vspace{-4mm}
    \footnotesize
$^a$for $T \leq 300{\rm K}$\\
$^b$for $T > 300{\rm K}$
\end{table}

Table \ref{tab:O_sources_reaction} includes all oxygen producing reactions -- besides photodissociation and the reactions listed in Tables 1-5 -- which are considered in this study.
We note that the second reaction in Table \ref{tab:O_sources_reaction}, i.e. \ce{C(^3P) + O2} mainly produces excited \ce{O(^1D)} as measured by \cite{ogryzlo_vibrational_1973} and \cite{costes_state_1998}, although some portion of ground state \ce{O(^3P)} cannot be ruled out. For the present purpose we assume all oxygen products of this reaction beeing
excited.

The change of enthalpy $\Delta H$ which is equal to the excess energy and listed in Table \ref{tab:O_sources_reaction} is obtained via Hess's law. According to this law, the change of enthalpy for a chemical reaction can be calculated by means of the difference $\Delta H_{\rm f}^\circ$ between the heat of formation of the products and the heat of formation of the reactants, i.e.
\begin{equation}
  \Delta H = \sum_{products} \Delta H_{\rm f}^\circ - \sum_{reactants} \Delta H_{\rm f}^\circ .
\end{equation}
If $\Delta H$ is negative, the reaction is exothermic, otherwise it is endothermic. The values for the heat of formation $\Delta H_{\rm f}^\circ$ of the products and reactants are taken from the \textit{CCCBDB (Computational Chemistry Comparison and Benchmark Database)} website\footnote{NIST Computational Chemistry Comparison and Benchmark Database.
NIST Standard Reference Database Number 101 Release 15b, August 2011; Editor: Russell D. Johnson III; http://cccbdb.nist.gov/}.

Figure 1 also includes the production rates according to the reactions listed in Table \ref{tab:O_sources_reaction} for which the maximum value is at least $1\,{\rm cm}^{-3}\,{\rm s}^{-1}$.

\subsection{Main Sources of Hot C Atoms}
\subsubsection{Dissociative and Radiative Recombination}
%-------------------------------------------------------
Dissociative recombination of \ce{CO2+} proceeds through the channels listed in Table 2. Atomic carbon is only produced via
the second and fourth channel, however, since the fourth channel represents an endothermic process only the second
channel is taken into account. The probability for this channel is low, however, because, as noted by \cite{viggiano_rate_2005},
it requires an unlikely rearrangement of the two oxygen atoms. The \ce{CO2+} ion is a linear molecule \ce{O=C=O+} where the
carbon is located in between the two oxygen atoms. In order to form the \ce{O2} molecule, the two oxygen atoms have to bond in a very short time.
It should be noted, however, that according to \cite{seiersen_dissociative_2003} the branching ratio of the second channel is 9\%,
in contrast to almost zero percent as reported by \cite{viggiano_rate_2005}.

Dissociative recombination of CO$^+$ takes place via the channels shown in Table 4. C atoms can be produced in the ground state C($^3$P)
and in the excited states C($^1$D) and C($^1$S). The latter states have energies of $1.26\,{\rm eV}$ and $2.68\,{\rm eV}$, respectively,
above the ground state.

\begin{table}[b!]
%\small
\centering
\caption{Possible sources of hot C atoms due to dissociative and radiative recombination and references of the rate coefficients.}\vspace{2mm}
\begin{tabular}{@{} p{4.5cm}  p{6.5cm} @{}}
\hline
     {\bf Source}     &   {\bf Rate coefficient [$\mathbf{cm^3\,s^{-1}}$]}   \\
\hline
	\multicolumn{2}{c}{\bf Dissociative recombination}   \\
	\ce{CO2+ + e -> C + O2}  &   \cite{viggiano_rate_2005}     \\
	                         &   $\quad \alpha=4.20\times10^{-7}$, $\beta=-0.75$  \\ %T > 1200\,{\rm K}:
	\ce{CO+ + e -> C + O}   &   \cite{rosen_absolute_1998}     \\
	                         &   $\quad \alpha=2.75\times10^{-7}$, $\beta=-0.55$  \\ %T \leq 1000\,{\rm K}:
	\multicolumn{2}{c}{\bf Radiative recombination}   \\
	\ce{C+ + e -> C + h$\nu$}   &   \cite{nahar_electronion_1997}  \\
	                         &   $\quad T \leq 7950\,{\rm K}: \alpha=4.67\times10^{-12}$, $\beta=-0.60$  \\
\hline
\end{tabular}
\label{tab:C_sources_recombination}
\end{table}

Radiative recombination of \ce{C+} proceeds via the reaction
\begin{align*}
	\ce{C+ + e -> C(^3P) + $h\nu$},
\end{align*}
where the parameters of equation (2) are taken from the UDFA06 database \citep{woodall_umist_2007} based on \cite{nahar_electronion_1997}. As for O$^+$, the production rates for radiative recombination of \ce{C+} are low, having a maximum of $10^{-5}\, {\rm cm}^{-3}\,{\rm s}^{-1}$ at $150\,{\rm km}$ for high solar activity and of $3 \times 10^{-6}\, {\rm cm}^{-3}\,{\rm s}^{-1}$ at $140\,{\rm km}$ for low solar conditions.

The coefficients $\alpha$ and $\beta$ for dissociative and radiative recombination are given in Table \ref{tab:C_sources_recombination}. The correponding atomic carbon production rates are illustrated in Figure 2 for low and high solar activity.

\begin{figure}[t]
	\begin{center}
	\includegraphics[width=0.65\columnwidth]{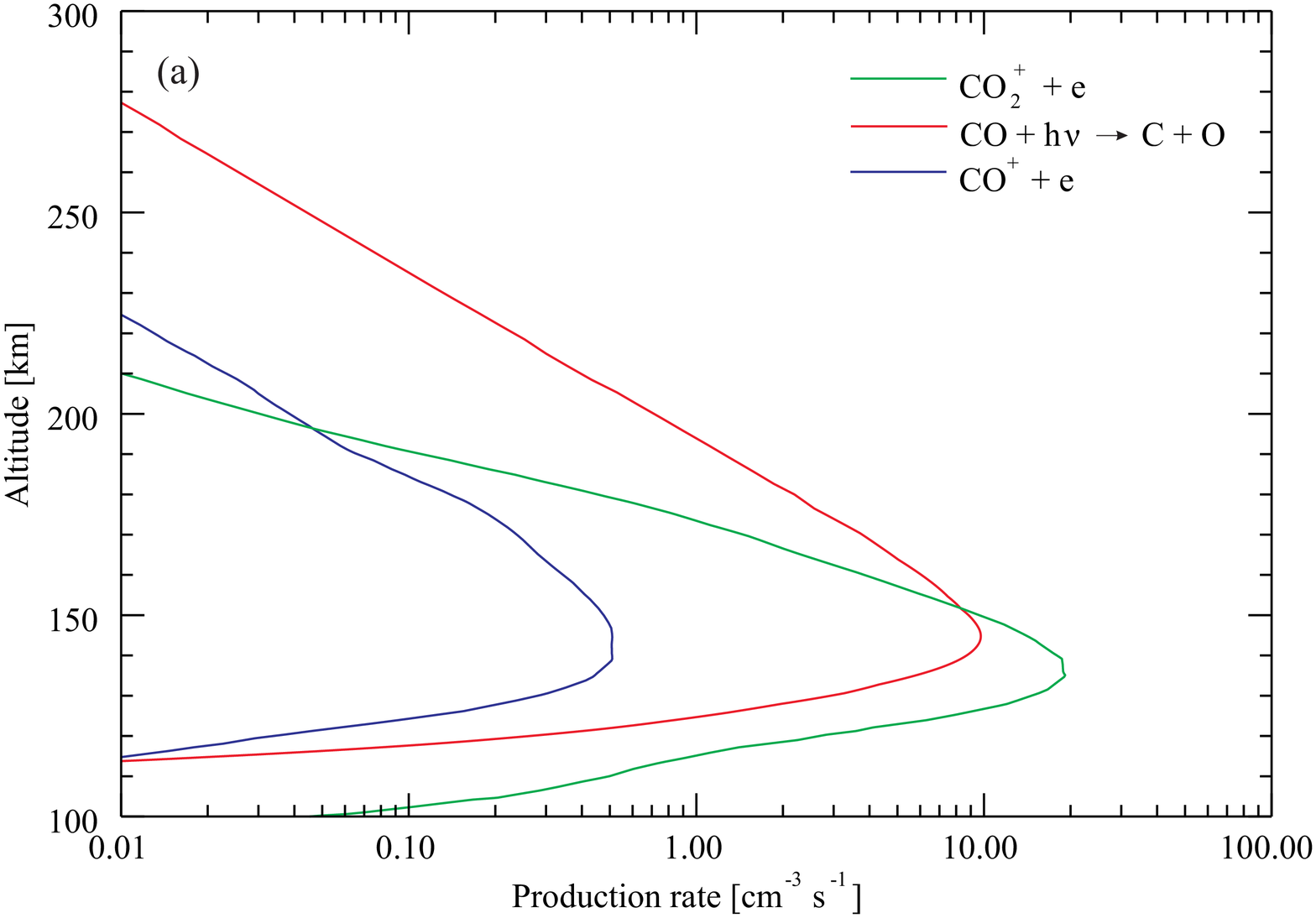}\vspace{4mm}
	\includegraphics[width=0.65\columnwidth]{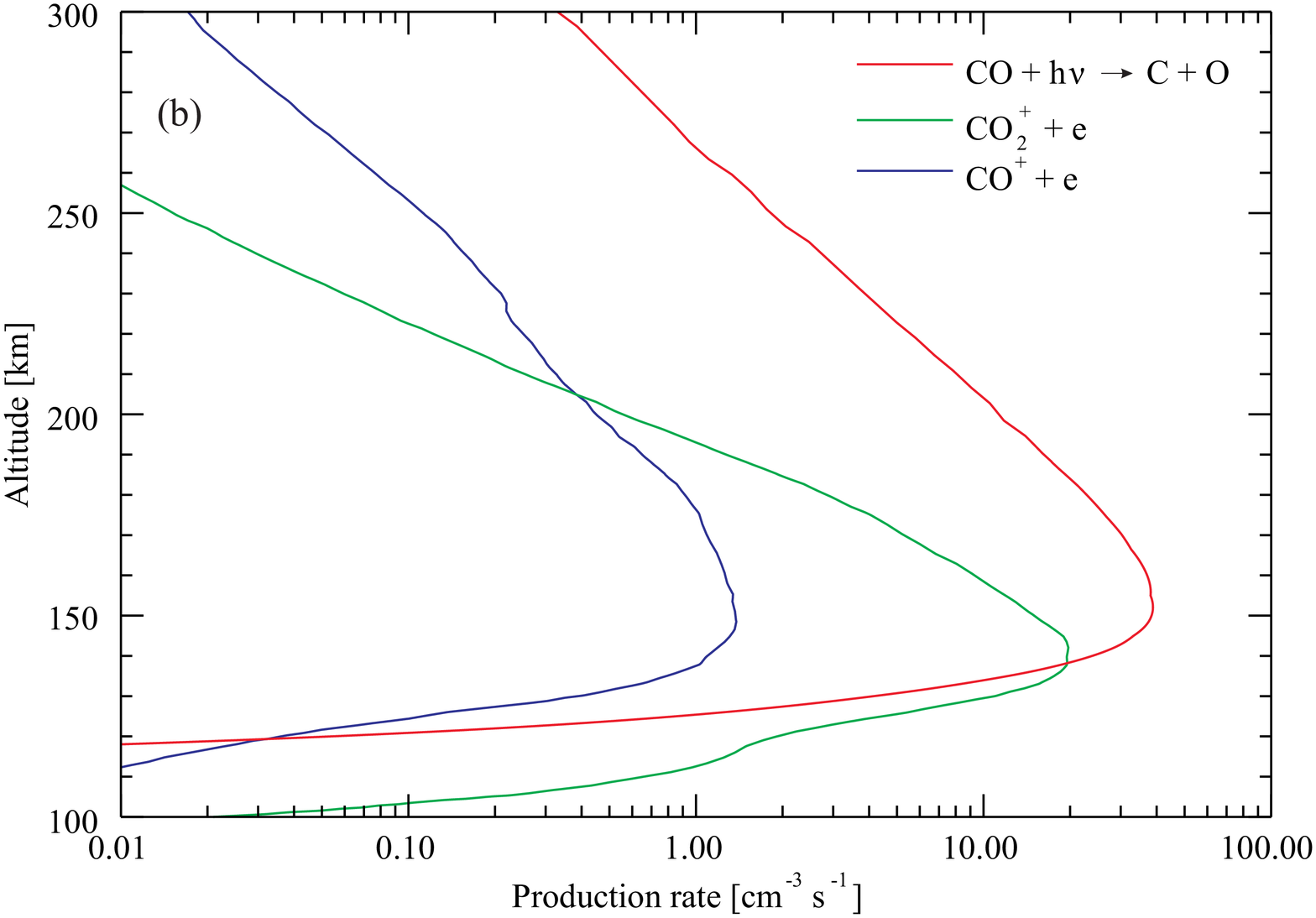}
	\caption{Production rates of hot C atoms for low (a) and high (b) solar activity due to different sources of carbon.}
	\label{fig:production_rates_C}
	\end{center}
\end{figure}

\subsubsection{Photodissociation}
In addition to dissociative recombination, photodissociation of \ce{CO} is another important source of hot C atoms
\begin{align*}
	\ce{CO + $h\nu$ -> C(^3\rm{P}) + O(^3\rm{P})}
\end{align*}
with a threshold energy of $11.14\,{\rm eV}$ (see \textit{Bond Dissociation Energies in Simple Molecules} - B. deB. Darwent. NSRDS-NBS 31,
(1970))\footnote{http://www.nist.gov/data/nsrds/NSRDS-NBS31.pdf}.
The production rate profiles of atomic carbon due to photodissociation of \ce{CO} are included in Figure 2.

\subsubsection{Chemical Reaction and Charge Transfer}
The coefficients $\alpha$, $\beta$, and $\gamma$ for some additional reaction that produce hot atomic carbon are listed
in Table \ref{tab:C_sources_reaction}. The corresponding production rates for high and low solar activity are also given,
showing that hot carbon is created by these reactions with a low rate only; therefore they are not displayed in Figure 2.

\begin{table}[b]
    \small
    \centering
    \caption{Possible sources of hot C atoms and the corresponding coefficients $\alpha$, $\beta$, and $\gamma$ to calculate the rate coefficient $k$ in ${\rm cm}^{3}\,{\rm s}^{-1}$. The excess energies for chemical reactions are given for the ground state of the products, e.g. \ce{C($^3\rm{P}$)}, \ce{CO(X$^1\Sigma^+$)}.}\vspace{2mm}
    \begin{tabular}{lccccc}
    \hline
    {\bf Source}     &   \multicolumn{3}{c}{\bf Rate coefficient [$\mathbf{cm^3\,s^{-1}}$]}   &  \multicolumn{2}{l}{\bf Production rate [$\mathbf{cm^3\,s^{-1}}$]} \\
         &  $\mathbf{\alpha}$   &  $\mathbf{\beta}$  &  $\mathbf{\gamma}$          &     {\bf HSA}    &     {\bf LSA}   \\
    \hline
	\multicolumn{4}{c}{\bf Chemical reaction}   \\
	\ce{C+ + NO -> NO+ + C + $1.99\,{\rm eV}$}   &   $5.20\times10^{-10}$ & $0.0$ & $0.0$     &   $< 6 \times 10^{-3}$   &   $< 4 \times 10^{-3}$\\
	\ce{He+ + CO2 -> O2+ + C + He + $1.04\,{\rm eV}$}   &   $1.10\times10^{-11}$ & $0.0$ & $0.0$   &   $< 10^{-4}$   &   $< 2 \times 10^{-4}$\\
	\ce{N+ + CO -> NO+ + C + $0.67\,{\rm eV}$}   &   $1.45\times10^{-10}$ & $0.0$ & $0.0$   &   $< 3 \times 10^{-2}$   &   $< 10^{-2}$\\
    \hline
    \end{tabular}
    \label{tab:C_sources_reaction}
\end{table}

\subsection{Total and Differential Cross Sections}
%-------------------------------------------------
The motion of the hot particles through the thermosphere is dominated by collisions with the background atmosphere.
Thus the appropriate treatment of collisions is a crucial task in the simulation of hot coronae. Depending on the
collision model, e.g. a hard sphere approximation or energy dependent cross sections, a wide range of outcome can be
obtained \citep[e.g.][]{lichtenegger_elusive_2009, groller_venus_2010}. In the present paper, collisions are modeled
by using total and differential cross sections.

\begin{table}[b!]  %[thb!]
\centering
\setlength{\belowcaptionskip}{7pt}
\caption{Total and differential collision cross sections ($\sigma$,  $d\sigma$) for elastic and quenching collisions between O atoms together with the available energy ranges.}
\label{tab:collision_cs_O_O}
\begin{tabular}{ lll }
\hline
     \textbf{Hot Particle}    &   \textbf{Collision Type}    &    \textbf{References}           \\
    \hline %\midrule
    \ce{O(^3P)} & elastic & \cite{tully_low_2001} \\
          &       & \hspace{10px}$\sigma$: $1 - 10\,{\rm eV}$ \textit{(Figure 4)} \\
          &       & \hspace{10px}$d\sigma$: 1, 5, and $10\,{\rm eV}$ \textit{(Figure 2)} \\
    \hline %\midrule
    \ce{O(^1D)} & elastic & \cite{yee_energy_1987} \\
          &       & \hspace{10px}$\sigma$: $0 - 5\,{\rm eV}$ \textit{(Figure 4)}  \\
          &       & \hspace{10px}$d\sigma$: 0.1 and $1.0\,{\rm eV}$ \textit{(Figure 2b and 3b)}   \\
          & quenching & \cite{yee_collisional_1990} \\
          &       & \hspace{10px}$\sigma$: $0.001 - 5\,{\rm eV}$ \textit{(Figure 4)} \\
          &       & \hspace{10px}$d\sigma$: like \ce{O(^1D)} elastic \\
    \hline %\midrule
    \ce{O(^1S)} & elastic & \cite{yee_energy_1985}  \\
          &       & \hspace{10px}$\sigma$: $0 - 5\,{\rm eV}$ \textit{(Figure 4)}   \\
          &       & \hspace{10px}$d\sigma$: 1 and $5\,{\rm eV}$ \textit{(Figure 2b and 3b)} \\
          & quenching & \hspace{10px}$\sigma$: like \ce{O(^1D)} elastic  \\
          &           & \hspace{10px}$d\sigma$: like \ce{O(^1D)} elastic  \\
\hline %\bottomrule
\end{tabular}
\end{table}

Collisions between energetic particles and neutral background constituents can be either elastic, inelastic, or quenching. In elastic collisions the kinetic energy is conserved, only the scattering angle can change. During inelastic collisions kinetic energy can be converted into vibrational energy and into electronic excitation, the latter being usually less likely. Quenching collisions describe the de-excitation of the reactant by collision, i.e. the conversion of the internal energy of the reactant into kinetic energy.

The references for the used total and differential collision cross sections ($\sigma$ and d$\sigma$) as well as the fraction of energy ($f_{\rm E}$) which goes into internal energy can be found in Tables \ref{tab:collision_cs_O_O} and \ref{tab:collision_cs_O_N2}.
Inelastic collision of two oxygen atoms are not considered because the excitation energy in this case is higher than the energy released during the collision.

% ********************************************************
% ***                        O                         ***
% ********************************************************
\subsubsection{Collisions between \ce{O}($^3$P, $^1$D, $^1$S) and \ce{O}}
%------------------------------------------------------------------------
Collisions between energetic (hot) oxygen and a neutral thermal atom are either elastic
\begin{equation}
\ce{O_{\rm h}(^3P\mbox{, }^1D\mbox{, }^1S)} + \mbox{A}_{\rm th} \rightarrow \ce{O_{\rm h}(^3P\mbox{, }^1D\mbox{, }^1S)} + \mbox{A}_{\rm h},
\end{equation}
where $\mbox{A}_{\rm th}$ denotes the thermal background particles and O$_{\rm h}$ the hot oxygen, or quenching
\begin{equation}
  \ce{O_{\rm h}(^1D\mbox{, }^1S)} + \mbox{A}_{\rm th} \rightarrow \ce{O_{\rm h} (^3P)} + \mbox{A}_{\rm h} + \Delta E_{\rm q},
\end{equation}
with the released energy $\Delta E_{\rm q}$ being 1.97 eV and 4.19 eV for \ce{O(^1D)} and \ce{O(^1S)}, respectively.
\begin{table}[t!]  %[htb!]
\centering
\setlength{\belowcaptionskip}{7pt}
\caption{Total and differential collision cross sections ($\sigma$,  $d\sigma$) used in this study for elastic, inelastic, and quenching collisions between O and \ce{N2}, as well as the fraction $(f_{\rm E})$ of the kinetic energy which changes into internal energy of the colliding molecule together with the available energy ranges.}
\label{tab:collision_cs_O_N2}
\begin{tabular}{lll}  %{ p{2.6cm} p{3.0cm} p{8.3cm} }  %\toprule
    \hline
     \textbf{Hot Particle}    &   \textbf{Collision Type}    &    \textbf{References}           \\
    \hline %\midrule
    \ce{O(^3P)} & elastic & \cite{balakrishnan_slowing_1998} \\
          &       & \hspace{10px}$\sigma$: 0 - $3\,{\rm eV}$ \textit{(Figure 1)} \\
          &       & \cite{balakrishnan_quantum_1998}  \\
          &       & \hspace{10px}$d\sigma$: 1, 2, and $3\,{\rm eV}$ \textit{(Figure 3 to 5)} \\
          & inelastic &  \cite{balakrishnan_slowing_1998}  \\
          &       & \hspace{10px}$\sigma$: 0 - $3\,{\rm eV}$  \\
          &       & \hspace{10px}$d\sigma$: $3\,{\rm eV}$ \textit{(Figure 2)} \\
          &       & \hspace{10px}$f_{\rm E}$: 0 - $3\,{\rm eV}$ \textit{(Figure 8)} \\
    \hline %\midrule
    \ce{O(^1D)} & elastic & \cite{balakrishnan_translational_1999} \\
          &       & \hspace{10px}$\sigma$: 0 - $2\,{\rm eV}$ \textit{(Figure 1)}  \\
          &       & \hspace{10px}$d\sigma$: like \ce{O(^3P)}  \\
          & inelastic & \hspace{10px}$\sigma$: like \ce{O(^3P)}  \\
          &       & \hspace{10px}$d\sigma$: like \ce{O(^3P)} \\
          &       & \cite{tachikawa_translational_1997}  \\
          &       & \hspace{10px}$f_{\rm E}$: 0 - $\sim 0.8\,{\rm eV}$ \textit{(Table 1)} \\
          & quenching & \cite{matsumi_translational_1996} \\
          &       & \hspace{10px}$\sigma$: 0 - $\sim 0.9\,{\rm eV}$ \textit{(Figure 9)} \\
          &       & \hspace{10px}$d\sigma$: like \ce{O(^3P)} \\
          &       & \cite{zahr_theoretical_1975}  \\
          &       & \hspace{10px}$f_{\rm E}$: 0 - $1.2\,{\rm eV}$ \textit{(Table II)} \\
\hline %\midrule
    \ce{O(^1S)} & elastic, inelastic,  &   \\
                & and quenching        & like \ce{O(^1D)}   \\
\hline %\bottomrule
\end{tabular}
\end{table}

Since no data are available for the differential cross section of quenching collisions of excited \ce{O(^1D)} and \ce{O(^1S)} atoms, we assume them to be similar to the elastic collisions of \ce{O(^1D)} atoms. The cross sections simulated by \cite{yee_energy_1985}, \cite{yee_energy_1987}, \cite{yee_collisional_1990}, and \cite{tully_low_2001} for various types of collisions are illustrated in Figure 3a. Since hot particles may gain energies up to about $10\,{\rm eV}$, the values for the cross sections are extrapolated to this interval.

\begin{figure}[t]
  \centering
  \noindent\includegraphics[width=0.65\columnwidth]{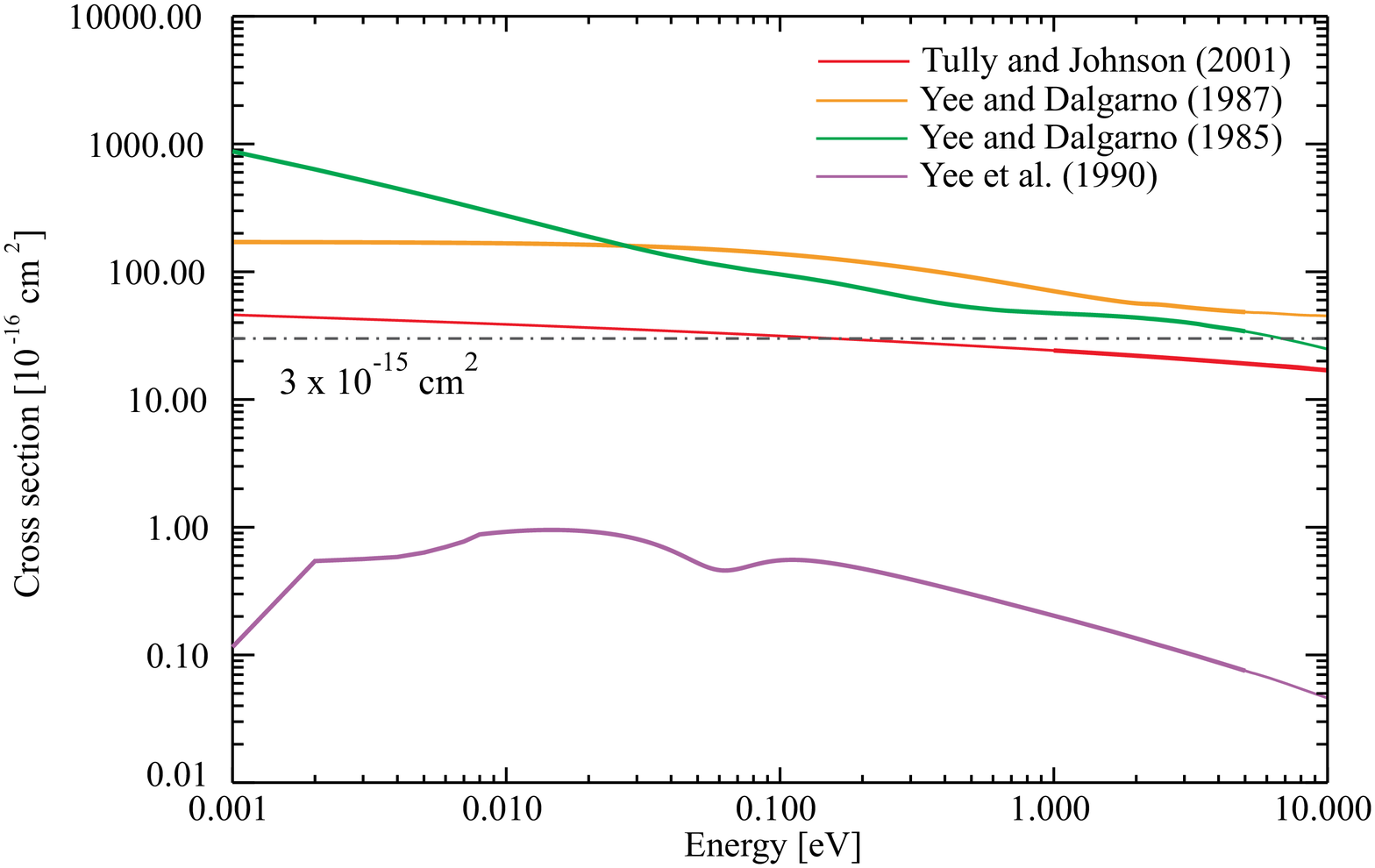}\vspace{5mm}
  \noindent\includegraphics[width=0.65\columnwidth]{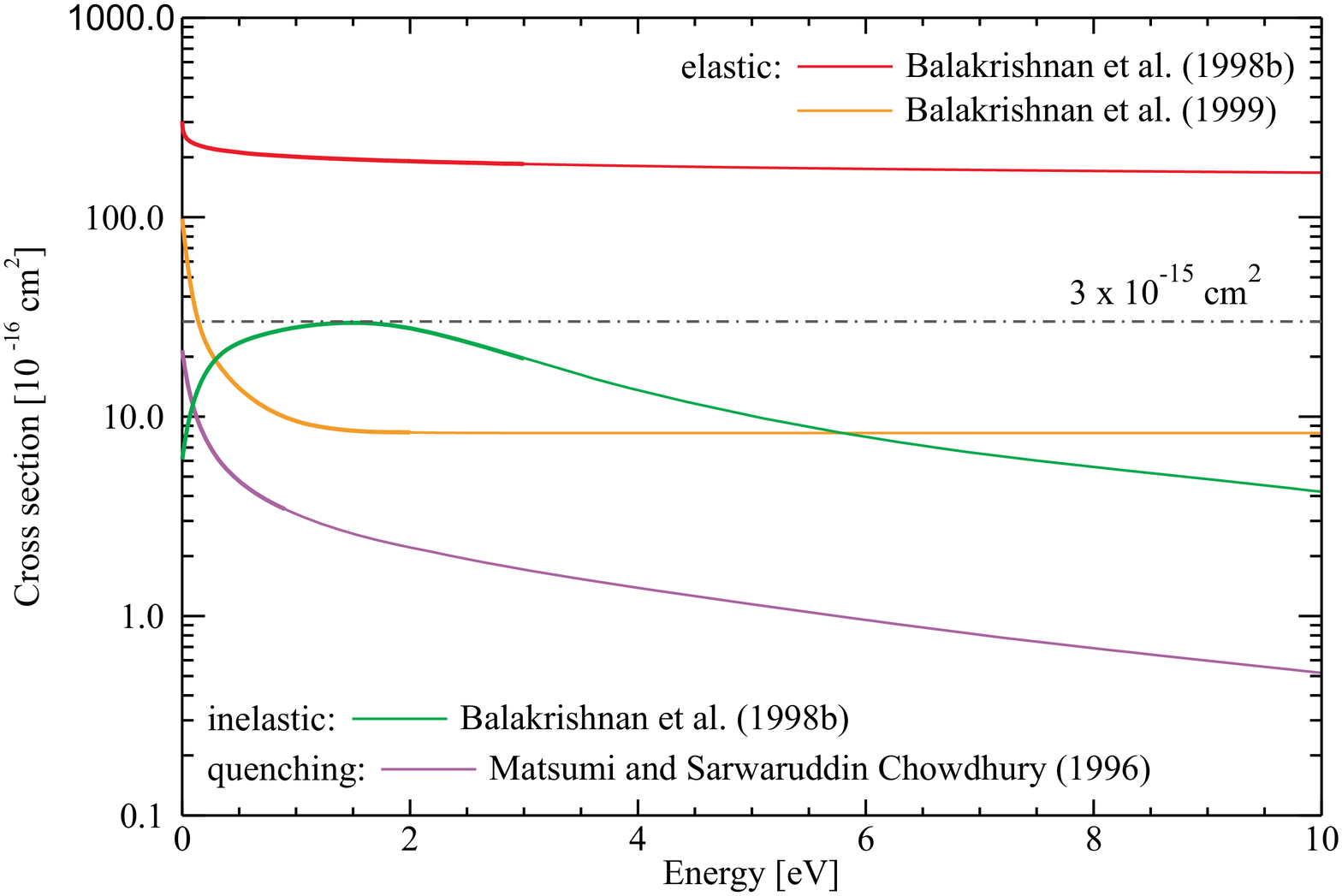}
  \caption{Total collision cross sections for (a) elastic and quenching collisions between \ce{O} atoms, (b) elastic, inelastic, and quenching collisions between \ce{O} atoms and \ce{N2} molecules. The indicated constant value of $3\times 10^{-15}\; \rm cm^2$ was used by some others in the past.}
  \label{fig:collision_cs_total}
\end{figure}

% ********************************************************
% ***                        N2                        ***
% ********************************************************
\subsubsection{Collisions between \ce{O(^3P\mbox{, }^1D\mbox{, }^1S)} and \ce{N2}}
%---------------------------------------------------------------------------------
Collisions between a suprathermal \ce{O} atom and a neutral atmospheric background molecule can be\\
\noindent
(a) elastic
\begin{equation}
  \ce{O_{\rm h}(^3P\mbox{, }^1D\mbox{, }^1S)} + \mbox{M}_{\rm th} \rightarrow \ce{O_{\rm h}(^3P\mbox{, }^1D\mbox{, }^1S)} + \mbox{M}_{\rm h},
\end{equation}
where $\mbox{M}_{\rm th}$ denotes the thermal background molecule and O$_{\rm h}$ the hot oxygen atom,\\
\noindent
(b) quenching
\begin{equation}
  \ce{O_{\rm h}(^1D\mbox{, }^1S)} + \mbox{M}_{\rm th} \rightarrow \ce{O_{\rm h}(^3P)} + \mbox{M}_{\rm h} + \Delta E_{\rm q},
\end{equation}
with the released energy $\Delta E_{\rm q}$ being a fraction of the excitation energies, or\\
\noindent
(c) inelastic
\begin{eqnarray}
    \ce{O_{\rm h}(^3P\mbox{, }^1D\mbox{, }^1S)} + \mbox{M}_{\rm th} \rightarrow \ce{O_{\rm h}(^3P\mbox{, }^1D\mbox{, }^1S)} + \mbox{M}_{\rm h} - \Delta E_{\rm i},
\end{eqnarray}
where $\Delta E_{\rm i}$ is the kinetic energy which is converted into internal energy of the products.

The total and differential collision cross sections for the collisions between O($^3$P, $^1$D, $^1$S) atoms and \ce{N2} molecules are taken from the references listed in
Table \ref{tab:collision_cs_O_N2}. In addition, energy ranges for which the cross sections are available are given together with the fraction $f_{\rm E}$ of the kinetic energy which
goes into internal energy of the colliding molecule.
The total collision cross section for inelastic \ce{O(^3P)-N2} collisions is calculated according to the fraction of the frequencies for elastic and inelastic collisions given in Figure 6 of \cite{balakrishnan_slowing_1998}. The fraction $f_{\rm E}$ of the energy which is deposited into N$_2$ vibrational and rotational excitation in percent of quenching \ce{O(^1D)-N2} collisions is obtained from Figure 3 and Equation 4.2 of \cite{zahr_theoretical_1975} for the three given collision energies of 0.03, 0.6, and $1.2\,{\rm eV}$.
For elastic collisions between \ce{O(^3P)} and \ce{N2} no differential cross sections are available, therefore the differential cross sections for \ce{N-N2} collisions published in \cite{balakrishnan_quantum_1998} (Figures 3 to 5) are used. The cross sections of \cite{zahr_theoretical_1975}, \cite{tachikawa_translational_1997}, \cite{balakrishnan_slowing_1998}, and \cite{balakrishnan_translational_1999} are based on simulations whereas those of \cite{matsumi_translational_1996} are obtained from measurements. The total collision cross sections are shown in Figure 3b together with the constant cross section used by e.g. \cite{fox_photochemical_2009}. Since the simulated particles can achieve energies up to about $10\,{\rm eV}$, the cross sections are extrapolated to this interval.

% ********************************************************
% ***              other molecules and ions            ***
% ********************************************************
\subsubsection{Collisions between \ce{O(^3P\mbox{, }^1D\mbox{, }^1S)} and other atoms and molecules}
%---------------------------------------------------------------------------------------------------
Due to the lack of data for collisions between hot oxygen and \ce{CO2} or CO, the same total and differential cross sections and energy loss rates as for \ce{O-N2} collisions are employed.
Collisions of hot oxygen with H and He are treated by means of the elastic collision cross section for \ce{O-H} collisions given in \cite{Zhang_JGR_2009}.
For the collisions of hot oxygen atoms with other background atoms, the data for \ce{O-O} collisions are used.

% ********************************************************
% ***                 other elements (C)               ***
% ********************************************************
\subsubsection{Collisions between \ce{C(^3P\mbox{, }^1D\mbox{, }^1S)} and atoms or molecules}
%--------------------------------------------------------------------------------------------
Since no data for collisions between hot atomic carbon and atoms or molecules of the background atmosphere are available, the same total and differential cross sections and energy loss rates as for collisions of hot oxygen with atoms and molecules are assumed, see Tables \ref{tab:collision_cs_O_O} and \ref{tab:collision_cs_O_N2}.

\section{Results and Discussion}
%-------------------------------
According to the input parameters used in our simulations, the calculated exobase altitude for low and high solar activity at Mars is located at about 220 and $260\,{\rm km}$, respectively, above the surface.
Once the energy distribution functions of hot O and C atoms at the exobase, at $380\,{\rm km}$ and at $450\,{\rm km}$ altitude are determined, escape fluxes as well as density profiles of the suprathermal atoms can be estimated.
Assuming that the SZA=60$^\circ$ eroded models represent an average over the martian dayside, global escape rates are obtained
by integrating the escape fluxes based on the according EDF and by multiplying the resulting total escape flux by the hemispheric area of Mars.

\begin{table}[b!]
\renewcommand*\arraystretch{0.8}
    \setlength{\belowcaptionskip}{7pt}
    \caption{Escape fluxes and hemispheric loss rates of hot oxygen atoms according to the present study. The numbers are based on the EDFs at the exobase altitude, 380 km and 450 km altitude, respectively.}
    \label{tab:O_escape}
    \centering
    \begin{tabular}{lcccccc}
    \hline
    {\bf Source}     &   \multicolumn{3}{c}{\bf Escape flux [$\mathbf{{\bf cm}^{-2}\,{\bf s}^{-1}}$]}    &    \multicolumn{3}{c}{\bf Escape rate [$\,\mathbf{{\bf s}^{-1}}$]}     \\
    \hline
    &   \multicolumn{6}{c}{\bf Low solar activity} \\
    &   {\bf 220 km}   &   {\bf 380 km}   &   {\bf 450 km}  &   {\bf 220 km}   &   {\bf 380 km}  &   {\bf 450 km}  \\
    \ce{O2+ + e -> O + O}   &   $1.8 \times 10^7$  &   $1.4 \times 10^7$  &  $1.3 \times 10^7$   &   $1.5 \times 10^{25}$   &   $1.2 \times 10^{25}$   &   $1.2 \times 10^{25}$  \\
    \ce{CO2+ + e -> CO + O}   &   $1.5 \times 10^7$  &   $1.2 \times 10^7$   &   $1.1 \times 10^7$   &   $1.3 \times 10^{25}$   &   $1.0 \times 10^{25}$   &   $1.0 \times 10^{25}$   \\
    \ce{O2 + C -> CO + O(^1D)}   &   $8.4 \times 10^5$   & $6.1 \times 10^5$    &   $5.8 \times 10^5$   &   $6.9 \times 10^{23}$   & $5.5 \times 10^{23}$   &   $5.3 \times 10^{23}$    \\
    \ce{O2+ + N -> NO + O}   &   $1.0 \times 10^5$  &  $0.7 \times 10^5$   &   $0.7 \times 10^5$    &   $0.8 \times 10^{23}$    &  $0.6 \times 10^{23}$     &   $0.6 \times 10^{23}$   \\
    {\bf total}   &   $\mathbf{3.4 \times 10^7}$   &   $\mathbf{2.7 \times 10^7}$   &   $\mathbf{2.5 \times 10^7}$   &   $\mathbf{2.9 \times 10^{25}}$    &   $\mathbf{2.3 \times 10^{25}}$     &   $\mathbf{2.3 \times 10^{25}}$  \\
    \\
    &   \multicolumn{6}{c}{\bf High solar activity}   \\
    &   {\bf 260 km}   &   {\bf 380 km}   &   {\bf 450 km}    &   {\bf 260 km}   &   {\bf 380 km}   &   {\bf 450 km}\\
    \ce{O2+ + e -> O + O}   &  $2.5 \times 10^7$   &   $2.1 \times 10^7$    &   $2.0 \times 10^7$   &   $2.1 \times 10^{25}$   &   $1.9 \times 10^{25}$     &    $1.9 \times 10^{25}$   \\
    \ce{CO2+ + e -> CO + O}   &   $1.3 \times 10^7$   &   $1.0 \times 10^7$   &  $1.0 \times 10^7$   &   $1.1 \times 10^{25}$   &   $0.9 \times 10^{25}$      &   $0.9 \times 10^{25}$   \\
   \ce{O2+ + N -> NO + O}   &   $11.4 \times 10^5$   &   $9.2 \times 10^5$   &   $8.5 \times 10^5$    &  $9.6 \times 10^{23}$  &   $8.2 \times 10^{23}$     &    $7.8 \times 10^{23}$   \\
    \ce{O2 + C -> CO + O(^1D)}   &   $8.1 \times 10^5$   &  $5.9 \times 10^5$  &  $5.5 \times 10^5$   &   $6.8 \times 10^{23}$  &  $5.2 \times 10^{23}$   &   $5.1 \times 10^{23}$   \\
    {\bf total}   &   $\mathbf{4.0 \times 10^7}$   &   $\mathbf{3.3 \times 10^7}$  &  $\mathbf{3.1 \times 10^7}$   &  $\mathbf{3.4 \times 10^{25}}$   &   $\mathbf{2.9 \times 10^{25}}$   &   $\mathbf{2.9 \times 10^{25}}$  \\
    \hline
    \end{tabular}
\end{table}

\subsection{Hot O Atoms}
%------------------------
\begin{table} %[t!]
\begin{center}
\setlength{\belowcaptionskip}{7pt}
\caption{Comparison of hot oxygen escape rates of various studies due to DR of \ce{O2+} for low (LSA) and high solar activity (HSA).}
\begin{tabular}{lcc}
\hline
    {\bf References}     &   \multicolumn{2}{c}{\bf Escape rate $\mathbf{\times\, 10^{25}\; {\rm s}^{-1}}$}  \\
          &       {\bf LSA}   &   {\bf HSA}  \\
    \hline
   present study (at the exobase level)  &   1.5   &   2.1   \\
   present study (at $450\,{\rm km}$)  &   1.2   &   1.9   \\
   \cite{kim_solar_1998}$^a$  &  3.4   &  8.5   \\
   \cite{hodges_distributions_2000}$^b$  &  2.8   &   -  \\
   \cite{hodges_rate_2002}$^c$  &  4.4   &  18.0   \\
   \cite{krestyanikova_stochastic_2006}$^d$  &  4.5  &  -  \\
   \cite{cipriani_martian_2007}$^e$ &  3.4   &  8.5   \\
   \cite{cipriani_martian_2007}$^f$ &  0.55   &  2.6   \\
   \cite{chaufray_mars_2007}$^g$  &  1.0   &  4.0   \\
   \cite{fox_photochemical_2009}$^h$  &  0.66   &  0.68   \\
   \cite{fox_photochemical_2009}$^i$  &  14.4   &  21.0   \\
   \cite{valeille_study_2010}$^j$  &  3.8   &  13   \\
   \cite{yagi_icarus_2012}$^k$  &  2.9 - 5.3   &  -   \\
\hline
\end{tabular}
\label{tab:O_escape_comparison}
\end{center}
\footnotesize
$^a$corrected by a factor of 6.5 by \cite{nagy_hot_carbon_2001}; two-stream model and constant collision cross section of $1.2 \times 10^{-15}\,{\rm cm}^{2}$\\
$^b$constant collision cross section of $2 \times 10^{-15}\,{\rm cm}^{2}$ for \ce{O-O} and \ce{O-CO2}\\
$^c$constant collision cross section of $2 \times 10^{-15}\,{\rm cm}^{2}$ for \ce{O-O} and \ce{O-CO2} and including quenching of \ce{O(^1D)} and \ce{O(^1S)}\\
$^d$Model B: including elastic, inelastic and quenching collisions; collision cross sections are taken from \cite{balakrishnan_slowing_1998} and \cite{kharchenko_energy_2000}\\
$^e$atmosphere inputs from \cite{kim_solar_1998} (corrected values) and the collision cross sections taken from \cite{kharchenko_energy_2000} and \cite{tully_low_2001}\\
$^f$atmosphere inputs from \cite{krasnopolsky_mars_2002} and the collision cross sections taken from \cite{kharchenko_energy_2000} and \cite{tully_low_2001}\\
$^g$single interaction potential and only atomic background atmosphere\\
$^h$eroded ion profiles, exobase approximation, isotropic scattering and constant collision cross section of $3 \times 10^{-15}\,{\rm cm}^{2}$\\
$^i$eroded ion profiles, forward scattering using differential cross sections from \cite{kharchenko_energy_2000} and constant collision cross section of $3 \times 10^{-15}\,{\rm cm}^{2}$\\
$^j$3D thermospheric inputs from \cite{Bougher_seasonal_variations_2006}; constant collision cross section of $2\times 10^{-15}\,{\rm cm^2}$ for \ce{O-O} and \ce{O-CO2} collisions\\
$^k$universal potential model for collisions; variations due to different martian seasons 
\end{table}

The escape fluxes together with the escape rates for various sources of hot oxygen are listed in Table \ref{tab:O_escape} for LSA and HSA.
Since the escape rates are calculated from the EDF, they depend on the altitude at which the EDF is taken. At sufficiently high altitudes,
however, where collisions become insignificant, the escape rates converge towards a constant value. To illustrate this effect, the escape
rates based on the EDFs at three different altitudes are shown in Table \ref{tab:O_escape}.

Although the production rates are generally larger during HSA, there is little difference in the escape of hot O between HSA and LSA. This is mainly due to the fact that at HSA also the neutral background density is higher, leading to an increase in the number of collisions between the suprathermal particles and the cold atmosphere and thus to a more efficient thermalization of the hot particles compared to LSA conditions.
It should be noted that this effect is only found for the eroded ionosphere models. In case of the non-eroded ionosphere models of
\cite{fox_photochemical_2009} the higher production rates of hot O at HSA is not totally compensated by the more efficient thermalization in the denser atmosphere; indeed in that case the escape rate of hot oxygen produced from \ce{O2+} is about 65\% higher at HSA than at LSA.

Moreover, inspection of Table \ref{tab:O_escape} also reveals that the escape flux due to DR of \ce{CO2+} is smaller for HSA than for LSA.
This is due to the fact that the production rates in the region close to the exobase are comparable for LSA and HSA ($\sim$0.1 and $\sim$0.3 cm$^{-3}$ s$^{-1}$), while the initial energies of the newly produced hot O is different because of the different ion temperatures at LSA and HSA. The loss rates for both solar conditions are mainly sustained by particles with initial energies between $4.7\lesssim E\lesssim 5.8$ eV whose production rates are higher at LSA than at HSA. Thus, although the number of particles with $E>5.8$ eV is increased at HSA with respect to LSA, the number of particles with $E<5.8$ eV is decreased, leading to a net escape flux at HSA smaller than at LSA.

The same effect can also be seen for the \ce{O2 + C} reaction. In case of \ce{O2+ + N}, however, the escape flux for HSA is higher than for LSA. In this case, the effect of the initial energy distribution plays a minor
role with respect to the higher production rate at higher altitudes.

Table \ref{tab:O_escape} also illustrates the effect of ignoring the production of hot particles and their collisions
with the background atmosphere above the exobase. The loss rates are found to be higher when based on the EDF at the exobase
than when based on the EDF at 450 km altitude. This shows that the hot particles produced above the exobase only marginally contribute to the total escape despite the fact that most of these particles will be lost to space if their energy exceeds the escape energy. Rather, the decrease of escaping oxygen due to energy loss in collisions above the exobase somewhat exceeds the gain due to production, so that the escape rate is in fact reduced at 450 km by $\sim 20-30$\% with respect to the values at the exobase altitude. In general our simulation results suggest that collision effects are still noticeable up to $\sim 5-10$ scale heights above the exobase.

In Table \ref{tab:O_escape_comparison} our escape rates are compared to previous simulations. The differences in the results mainly reflect the differences in the input neutral and ion density profiles and the different treatment of collisions. Although we use
the neutral density and ionosphere model of \cite{fox_photochemical_2009}, the escape rates differ by a factor of $0.5-10$.
The main reason may be attributed to the different collision models: While \cite{fox_photochemical_2009} uses a constant collision
cross section of $3\times 10^{-15}\,\rm cm^2$ for all types of collisions, we invoke different energy dependent collision cross
sections (see Tables \ref{tab:collision_cs_O_O} and \ref{tab:collision_cs_O_N2}). In particular, the cross section for collisions with molecules -- which are the most important types of collisions up to $\sim 200\,{\rm km}$ -- is about one order of magnitude higher (see Figure 3b) than the above constant value. This higher value significantly increases the average number of collisions, thus leading to more efficient thermalization and to lower escape rates.
\begin{figure}[t]
	\begin{center}
	\includegraphics[width=0.65\columnwidth]{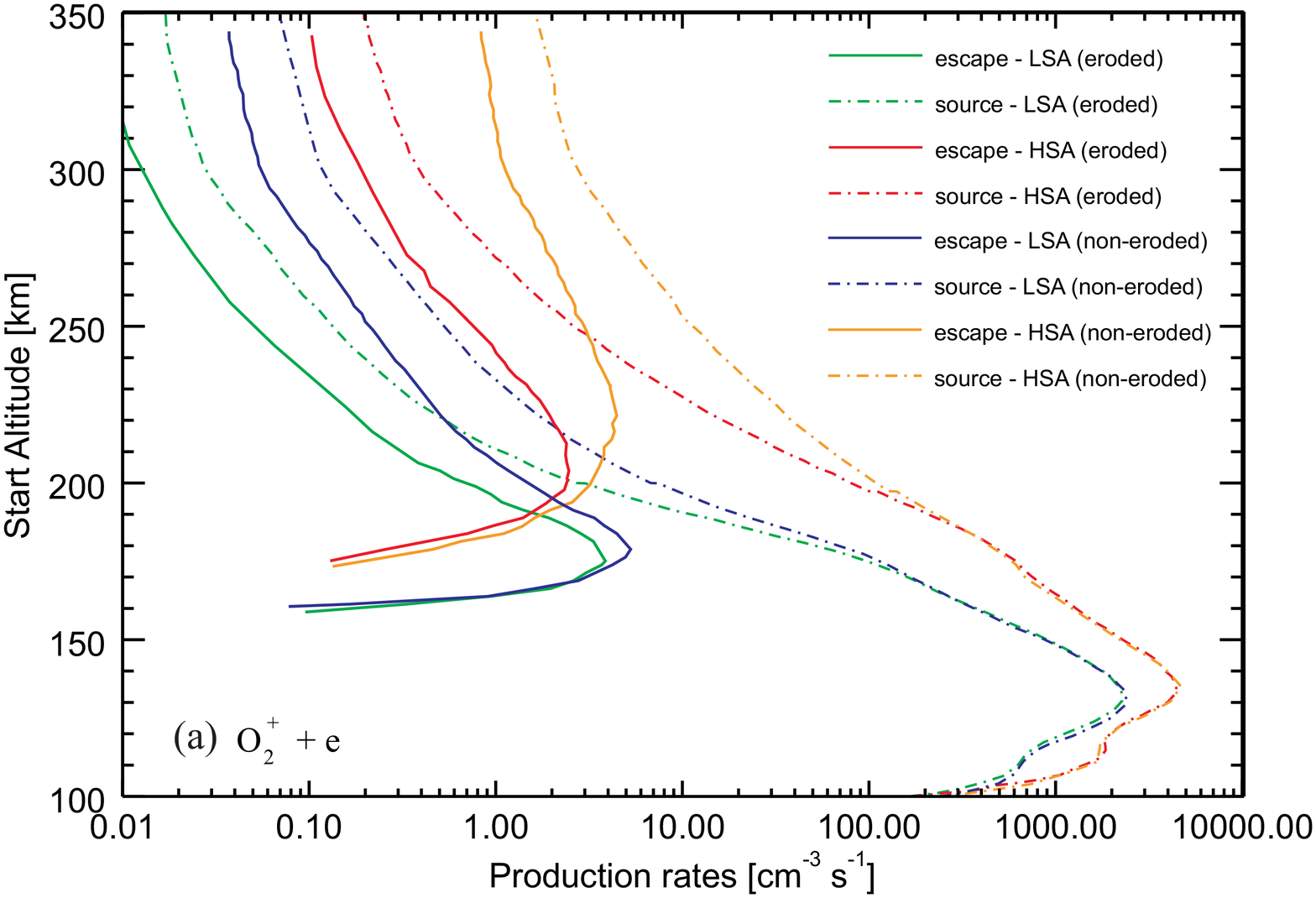}\vspace{5mm}
	\includegraphics[width=0.65\columnwidth]{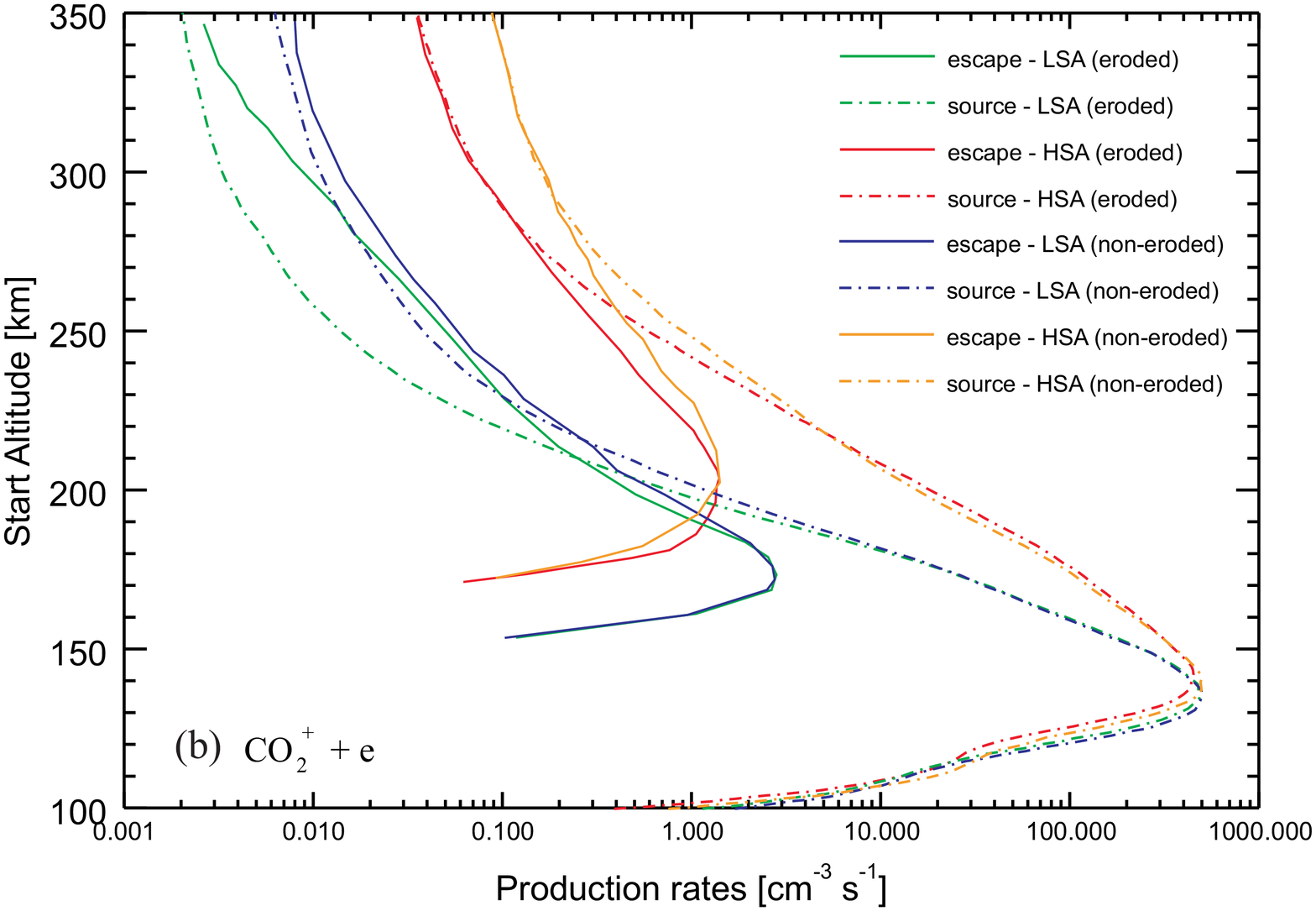}
	\caption{Production rates of escaping particles generated by DR of \ce{O2+} (a) and \ce{CO2+} (b) as a function of altitude
   for both solar activities and for the eroded and non-eroded ionosphere models. The dashed-dotted lines represent the production rates of all (escaping and non-escaping) oxygen atoms created via DR of \ce{O2+} (a) and \ce{CO2+} (b).}
    \label{fig:escape_rates_DR_of_O2plus}
	\end{center}
\end{figure}
Finally, while most previous studies reveal a distinct dependence of the hot oxygen loss rates on solar activity, this is neither the case in our results nor in those of \cite{fox_photochemical_2009} and may be attributed to the use of an eroded ionosphere model.
Figure 4 illustrates the production rates of escaping O resulting from the DR of \ce{O2+} and \ce{CO2+} as a function of altitude for both solar activities. For comparison, the results based on the eroded and non-eroded ionosphere models of \cite{fox_photochemical_2009} are shown. The solid lines represent the production rates at each height of those particles (consisting of hot particles generated by photochemical reactions and of background particles which became ''hot'' upon collisions with these particles)
which arrive at 380 km altitude with an energy exceeding the escape energy. As can be seen,
the production rates for the non-eroded models become larger than those for the eroded ones approximately at altitudes, where the
\ce{O2+} density profiles become sensitive to the upward velocity. At higher altitudes, the production rate of escaping hot O created by collisions can even exceed the local production rate of O due to DR of \ce{CO2+}.

The exosphere density profiles based on the energy distribution functions of hot oxygen at the exobase for low and high solar activity are shown in Figure 5.
\begin{figure}[t!]
	\begin{center}
	\includegraphics[width=0.65\columnwidth]{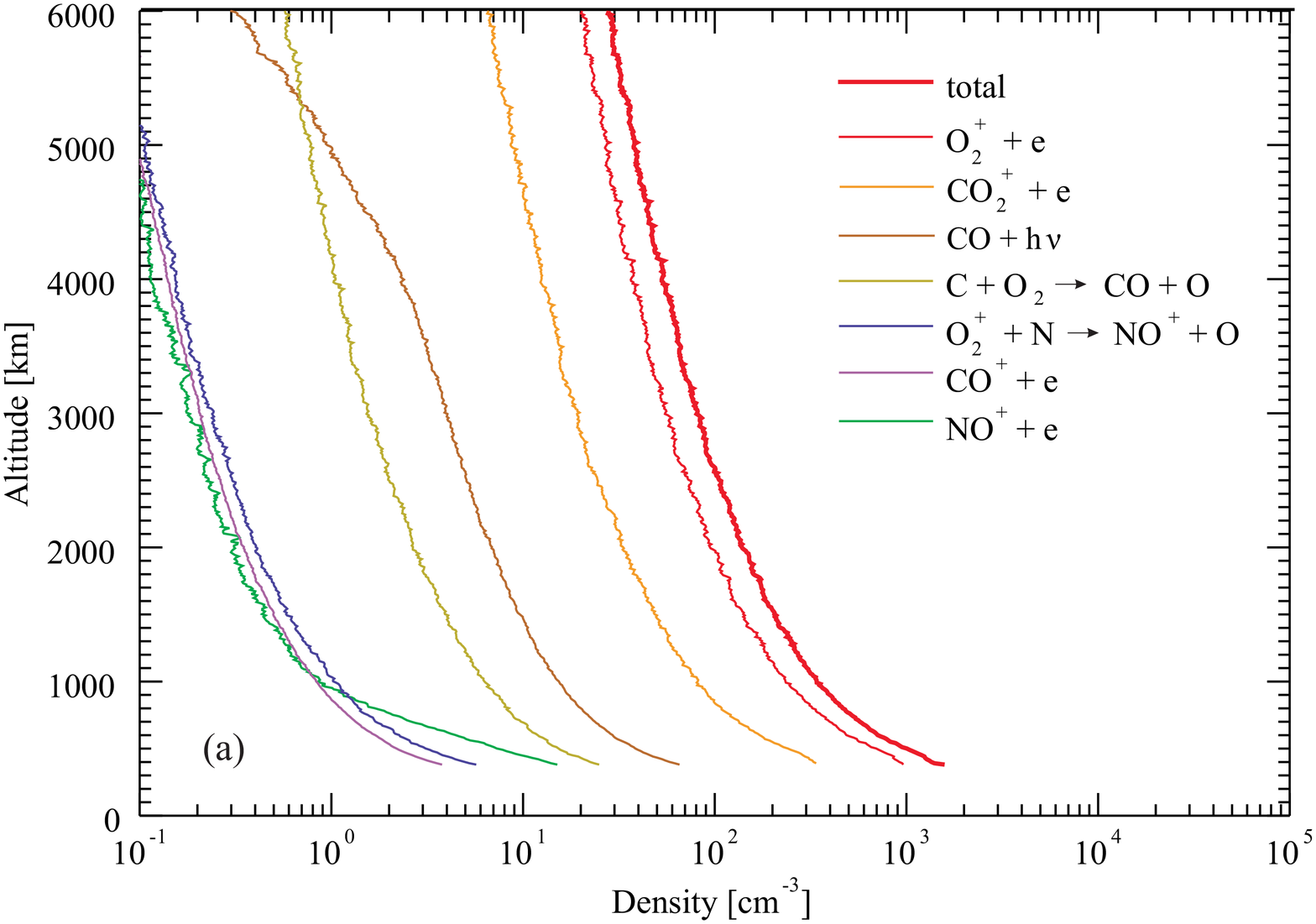}\vspace{5mm}
	\includegraphics[width=0.65\columnwidth]{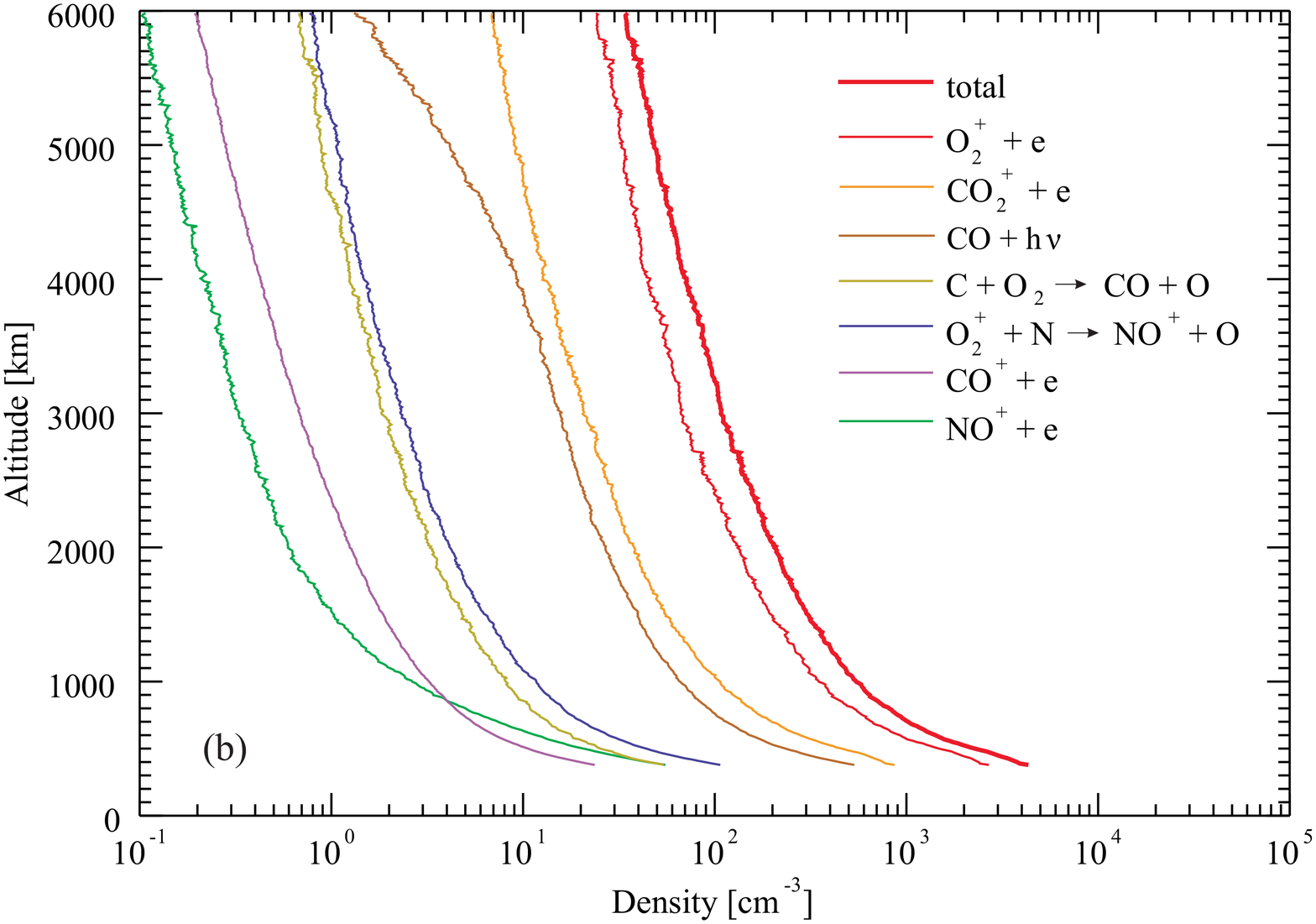}
	\caption{Exosphere densities of hot oxygen atoms for various sources for low (a) and high (b) solar activity.}
	\label{fig:density_O}
	\end{center}
\end{figure}
For both low and hight solar activity conditions the total exosphere density is mainly governed by DR of \ce{O2+} and \ce{CO2+}.
During HSA, however, at altitudes below $\sim 4000$ km, the oxygen density due to photodissociation of \ce{CO} becomes comparable with the density of O produced via DR of \ce{CO2+}.

It should be noted that, although the production rate of hot O due to \ce{NO+ + e} is between one and two orders of magnitudes higher than the one due to \ce{CO+ + e} (see Figure 1), the obtained exosphere densities are comparable. The main reason is that the initial energies of the O atoms originating from \ce{NO+ + e} are on average much lower than those originating from
\ce{CO+ + e}. Moreover, although the oxygen production rates due to the reactions \ce{C + O2} and \ce{NO+ + e} are comparable above $\sim130$ km (Figure 1), the corresponding exosphere density arising from \ce{C + O2} for HSA and LSA is nearly one order of magnitude higher than that arising from \ce{NO+ + e} (Figure 5).

\begin{figure}[t]
	\begin{center}
	\includegraphics[width=0.85\columnwidth]{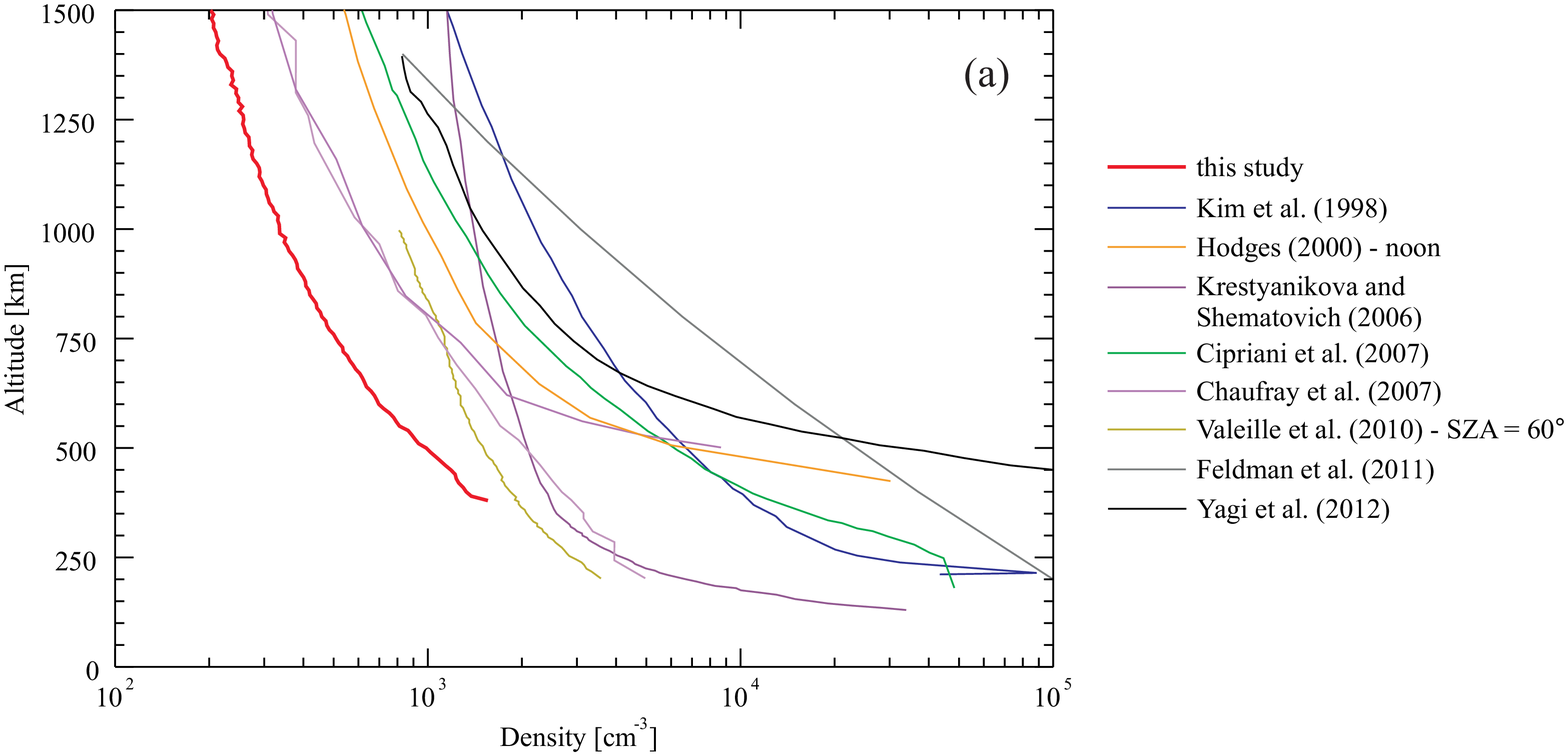}\vspace{5mm}
	\includegraphics[width=0.85\columnwidth]{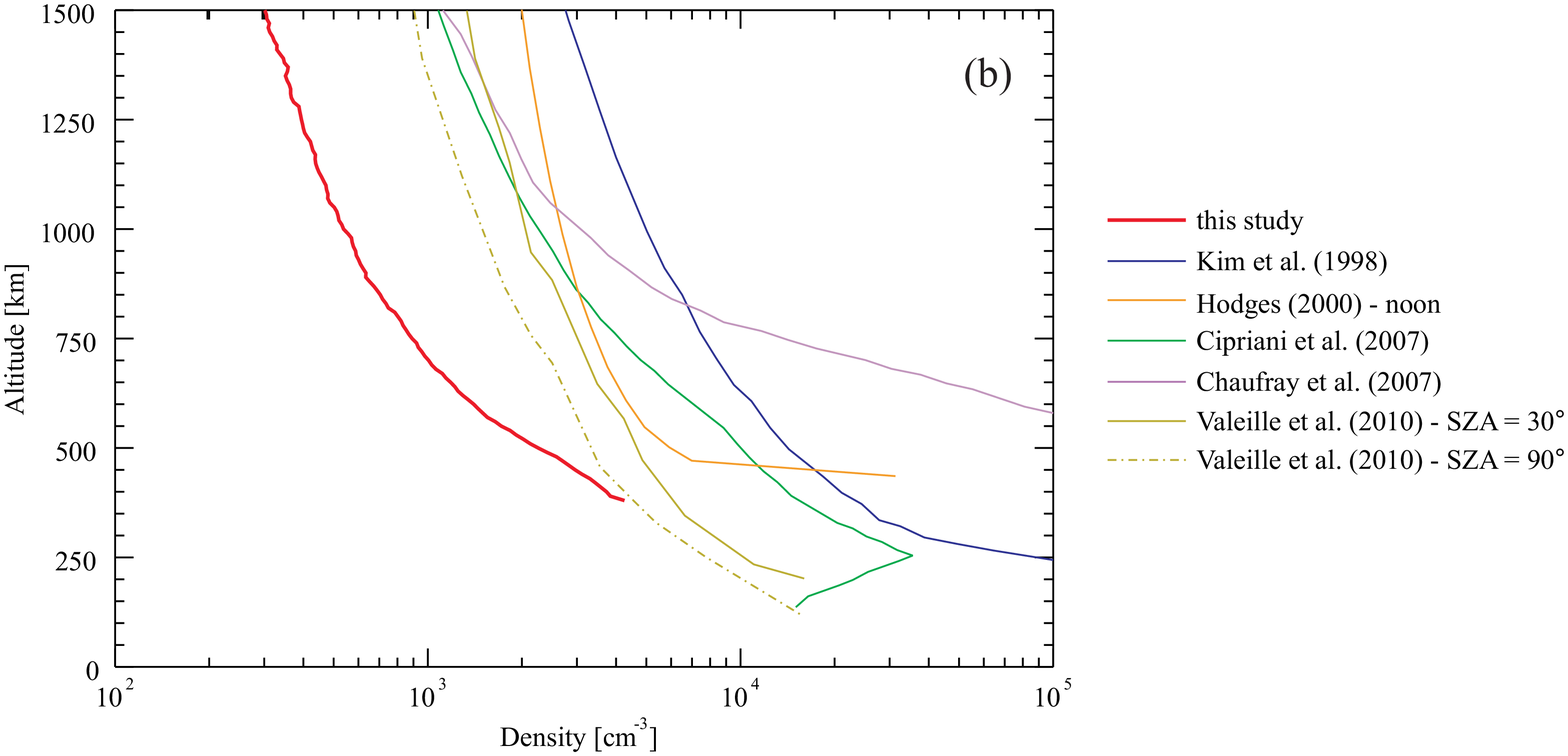}
	\caption{Comparison of exosphere densities of hot oxygen atoms for a number of studies. The curves of \cite{hodges_distributions_2000}, \cite{chaufray_mars_2007}, and \cite{yagi_icarus_2012} include also the cold component.}
	\label{fig:density_O_comparison}
	\end{center}
\end{figure}
This is due to the initial velocities of the oxygen atoms gained via DR of \ce{NO+}, which are much lower than those acquired via the reaction between C and \ce{O2}. Further, although the production rate of atomic oxygen originating from charge transfer between an ionized and a neutral O becomes dominant at higher altitudes (Figure 1), the contribution of this reaction to the exospheric density is negligible because the initial energies involved are similar to the average ion energies of $\sim 0.3$ eV.
For HSA as well as LSA the production rate of oxygen atoms with an excess energy of $4.17\,{\rm eV}$ due to reaction \ce{O2+ + N} is about three orders of magnitude smaller than the production rate due to DR of of \ce{O2+} and thus leading also to much lower exosphere densities.
Finally, at low solar activity, the oxygen exosphere density resulting from the reaction \ce{O2+ + N} is about 5-10 times smaller than the O density produced via \ce{C + O2} (Figure 5a) because above $\sim 130$ km the production rate is also smaller by a similar amount (Figure 1a). At high solar activity, however, the production rates of these two reactions above $\sim 180$ km become comparable (Figure 1b), making also the densities comparable (Figure 5b).

In Figure 6 our exosphere densities are compared with previous studies. This Figure reflects the variety of results obtained based on different assumptions. The use of energy dependent total and differential cross sections and the inclusion of elastic, inelastic and quenching collisions as done in this study (see Tables \ref{tab:collision_cs_O_O} and \ref{tab:collision_cs_O_N2}) appear to favour lower exosphere densities than other assumptions (Table \ref{tab:O_escape_comparison}). In addition, in our simulations the collision of ground-state oxygen atoms with the atmospheric background is modeled by means of the total and differential cross sections of \cite{tully_low_2001}\ (see Table \ref{tab:collision_cs_O_O}). As shown by \cite{groller_venus_2010} the use of the corresponding cross sections published by \cite{kharchenko_energy_2000} would lead to somewhat higher densities.

\subsection{Hot C Atoms}
%-----------------------
The escape fluxes as well as the escape rates for the most efficient sources of hot C are listed in Table \ref{tab:C_escape}.
As opposed to oxygen, where DR of \ce{O2+} and \ce{CO2+} are the most important suppliers for hot O escape, the C loss is
mainly due to photodissociation of CO.
As in the case of oxygen, the higher production rates at HSA are at least partially compensated by the simultaneously increased number of collisions with the denser background atmosphere,
leading to escape fluxes at HSA of about four times the flux at LSA. Regarding DR of \ce{CO2+}, the corresponding carbon escape flux at HSA is even lower than at LSA. It should also be noted that the carbon production and escape rate due to photodissociation of CO, i.e. of a neutral molecule, is not affected by the erosion of the ionosphere, in contrast to the production and escape of oxygen due to DR of \ce{O2+}- and \ce{CO2+}-ions.
Finally, the total loss rates calculated by means of the EDFs at 450 km altitude are about $\sim 25-30$\% smaller than the loss rates obtained by using the EDFs at the exobase level. Thus collisions above the exobase are again still sufficient to reduce the flux of escaping carbon by a noticeable amount.
\begin{table}[t!]
\renewcommand*\arraystretch{0.8}
    %\begin{center}
    \setlength{\belowcaptionskip}{7pt}
    \caption{Escape fluxes and hemispheric loss rates of hot carbon atoms according to the present study. The numbers are based on the EDFs at the exobase altitude, 380 km and 450 km altitude, respectively.}
%   \centering
    \begin{tabular}{lcccccc}
    \hline
    {\bf Source}     &   \multicolumn{3}{c}{\bf Escape flux [$\mathbf{{\bf cm}^{-2}\,{\bf s}^{-1}}$]}    & \multicolumn{3}{c}{\bf Escape rate [$\,\mathbf{{\bf s}^{-1}}$]}           \\
    \hline
    &   \multicolumn{6}{c}{\bf Low solar activity}   \\
    &   {\bf 220 km}   &   {\bf 380 km}   &   {\bf 450 km}    &   {\bf 220 km}   &   {\bf 380 km}   &   {\bf 450 km} \\
    \ce{CO + h$\nu$ -> C + O}     &   $10.6 \times 10^5 $    &   $7.8 \times 10^5 $   &   $7.1 \times 10^5 $   &   $8.7 \times 10^{23}$   &   $6.9 \times 10^{23}$   &   $6.5 \times 10^{23} $   \\
    \ce{CO2+ + e -> O2 + C}$^a$   &   $1.4 \times 10^5$   &   $1.0 \times 10^5$   &     $0.9 \times 10^5$    &   $1.1 \times 10^{23}$   &   $0.9 \times 10^{23}$   &   $0.8 \times 10^{23}$ \\
    \ce{CO+ + e -> C + O}         &   $0.8 \times 10^5$   &   $0.7 \times 10^5$   &   $0.6 \times 10^5$    &   $0.7 \times 10^{23}$   &   $0.6 \times 10^{23}$   &   $0.6 \times 10^{23}$   \\
    {\bf total} &   $\mathbf{12.8 \times 10^5}$   &   $\mathbf{9.5 \times 10^5}$    &   $\mathbf{8.6 \times 10^5}$     &   $\mathbf{10.5 \times 10^{23}}$   &   $\mathbf{8.4 \times 10^{23}}$   &   $\mathbf{7.9 \times 10^{23}}$   \\
    \\
    &    \multicolumn{6}{c}{\bf High solar activity}   \\
    &   {\bf 260 km}   &   {\bf 380 km}   &   {\bf 450 km}    &   {\bf 260 km}   &   {\bf 380 km}   &   {\bf 450 km}\\
    \ce{CO + h$\nu$ -> C + O}     &   $45.1  \times 10^5$    &   $34.4  \times 10^5$   &   $30.4 \times 10^5 $   &   $3.8  \times 10^{24}$   &   $3.1  \times 10^{24}$   &   $2.8 \times 10^{24} $   \\
    \ce{CO+ + e -> C + O}    &   $3.4 \times 10^5$    &   $3.3 \times 10^5$   &   $3.3 \times 10^5 $   &   $0.3  \times 10^{24}$   &   $0.3 \times 10^{24}$   &   $0.3 \times 10^{24} $   \\
    \ce{CO2+ + e -> O2 + C}$^a$     &   $1.2 \times 10^5$    &   $0.8 \times 10^5$     &   $0.8 \times 10^5$   &   $0.1 \times 10^{24}$   &   $0.1 \times 10^{24}$   &   $0.1 \times 10^{24}$    \\
    {\bf total} &   $\mathbf{49.7 \times 10^5}$  &   $\mathbf{38.5 \times 10^5}$   &    $\mathbf{34.5 \times 10^5}$    &   $\mathbf{4.2 \times 10^{24}}$   &   $\mathbf{3.5 \times 10^{24}}$   &    $\mathbf{3.2 \times 10^{24}}$    \\
    \hline
    \end{tabular}
    \label{tab:C_escape}
    %\end{center}
    \footnotesize\vspace{2mm}\\
    $^a$the branching ratio of this process is assumed to be 4\%\\
\end{table}

\begin{table}%[t!]
    \begin{center}
    \setlength{\belowcaptionskip}{7pt}
    \caption{Escape fluxes of hot atomic carbon obtained in previous studies.}
    \begin{tabular}{lcc}
    \hline
    {\bf Reaction}     &   \multicolumn{2}{c}{\bf Escape flux $\,\mathbf{\times\, 10^{5} {\bf cm}^{-2}\,{\bf s}^{-1}}$}  \\
          &       {\bf low solar activity}   &   {\bf high solar activity}  \\
    \hline
    \multicolumn{3}{c}{\bf \cite{nagy_hot_carbon_2001}$^a$}  \\
    \multicolumn{3}{c}{hemispheric escape fluxes} \\
    \ce{CO + h$\nu$ -> C + O}   &   -  &   60  \\
    \ce{CO+ + e -> C + O}       &   -  &   17  \\
    {\bf total}  &	 {\bf 5.33}   &   {\bf 77}  \\
    \multicolumn{3}{c}{\bf \cite{fox_photochemical_2001}$^b$} \\
    \multicolumn{3}{c}{global average escape fluxes$^c$}  \\
    \ce{CO + h$\nu$ -> C + O}   &   1.65   &   18  \\
    \ce{CO+ + e -> C + O}       &   0.29   &   6.2  \\
    {\bf total}			        &	{\bf 1.94}   &   {\bf 24.2}  \\
    {\bf total (including additional sources)} &	 {\bf 2.1}   &   {\bf 26}  \\
    \multicolumn{3}{c}{\bf \cite{fox_dissociative_2004}$^b$} \\
    \multicolumn{3}{c}{hemispheric escape fluxes}  \\
    \ce{CO + h$\nu$ -> C + O}   &   7.3   &   35  \\
    \ce{CO+ + e -> C + O}       &   1.28  &   7.7  \\
    \ce{CO2+ + e -> C + O2}     &   0.31  &   0.49 \\
    {\bf total}				 &	 {\bf 8.89}   &   {\bf 43.19}  \\
    {\bf total (including additional sources)} &	 {\bf 11.1}   &   {\bf 64}  \\
    \multicolumn{3}{c}{\bf \cite{cipriani_martian_2007}$^d$}  \\
    \multicolumn{3}{c}{hemispheric escape fluxes$^e$} \\
    \ce{CO+ + e -> C + O}     &   0.005   &   0.45  \\
    \hline
    \end{tabular}
    \label{tab:C_escape_comparison}
    \end{center}
    \footnotesize
    $^a$cross section of $1.2 \times 10^{-15}\,{\rm cm}^2$; for LSA only the sum of the rates for the two reactions is given by \cite{nagy_hot_carbon_2001}\\
    $^b$cross section of $3 \times 10^{-15}\,{\rm cm}^2$; exobase at $\sim 188\,{\rm km}$ for LSA and $\sim 220\,{\rm km}$ for HSA\\
    $^c$hemispheric escape fluxes are twice as large\\
    $^d$atmosphere inputs from \cite{kim_solar_1998} and the collision cross sections used from \cite{kharchenko_energy_2000} and \cite{tully_low_2001}\\
    $^e$calculated from the escape rates [s$^{-1}$] given in Table 1 of \cite{cipriani_martian_2007} by dividing through the martian hemispheric surface at $400\,{\rm km}$ altitude\\
\end{table}

In Table \ref{tab:C_escape_comparison} the escape rates of hot atomic carbon due to various source terms obtained in previous simulations are shown. As discussed in \cite{fox_dissociative_2004}, a detailed analysis of the differences among the escape fluxes obtained by \cite{nagy_hot_carbon_2001}, \cite{fox_photochemical_2001}, and \cite{fox_dissociative_2004} is difficult, but may be related to the use of different solar fluxes, dissociative recombination coefficients and exobase altitudes. The results of \cite{cipriani_martian_2007} are significantly lower than the aforementioned values; unfortunately, these authors do not discuss any possible reasons for this discrepancy.

Upon comparing our computed carbon escape fluxes (Table \ref{tab:C_escape}) with those published in \cite{fox_dissociative_2004}
(Table \ref{tab:C_escape_comparison}), it is seen that the fluxes at the exobase due to photodissociation of CO are higher by a factor
of $\sim 1.5$ and $\sim1.3$ for LSA and HSA, respectively.
Our escape fluxes at the exobase related to DR of \ce{CO2+} are about a factor of $\sim 4.5$ and $\sim 2.4$ higher for low and high solar activity, respectively, whereas our escape fluxes due to DR of \ce{CO+} are about a factor of $\sim 1.6$ and $\sim 2.3$ lower than the one given in \cite{fox_dissociative_2004} who assumes the exobase locations at 184 and 220 km altitude for low and high solar acitivity. Besides the differences in the exobase heights, these latter lower values can be explained mainly by our lower production rates (Figure 1) which are based on the \ce{CO+} density profiles of \cite{fox_photochemical_2009} being smaller than those given in \cite{fox_dissociative_2004}.

\begin{figure}[t]
	\begin{center}
	\includegraphics[width=0.65\columnwidth]{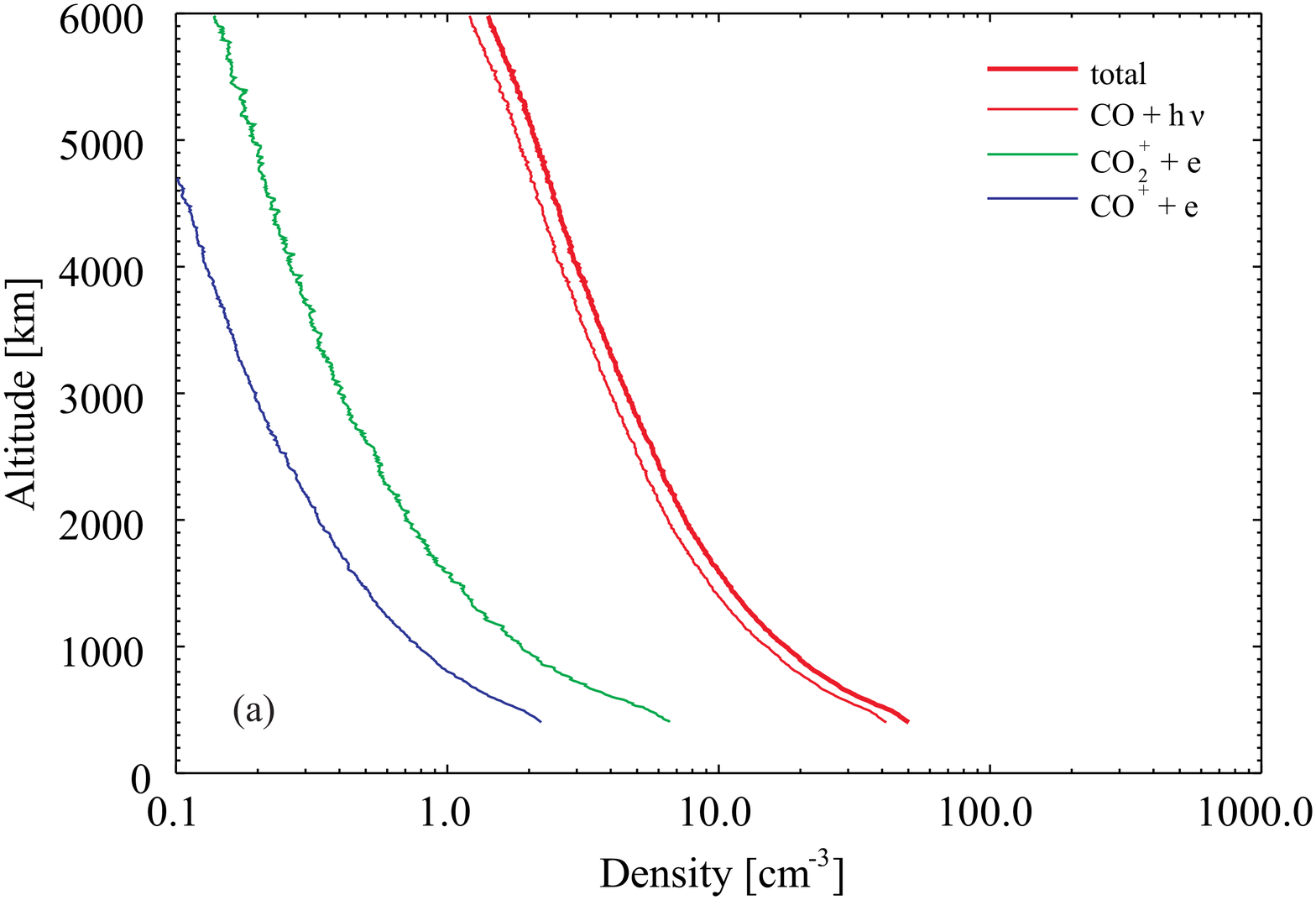}\vspace{5mm}
	\includegraphics[width=0.65\columnwidth]{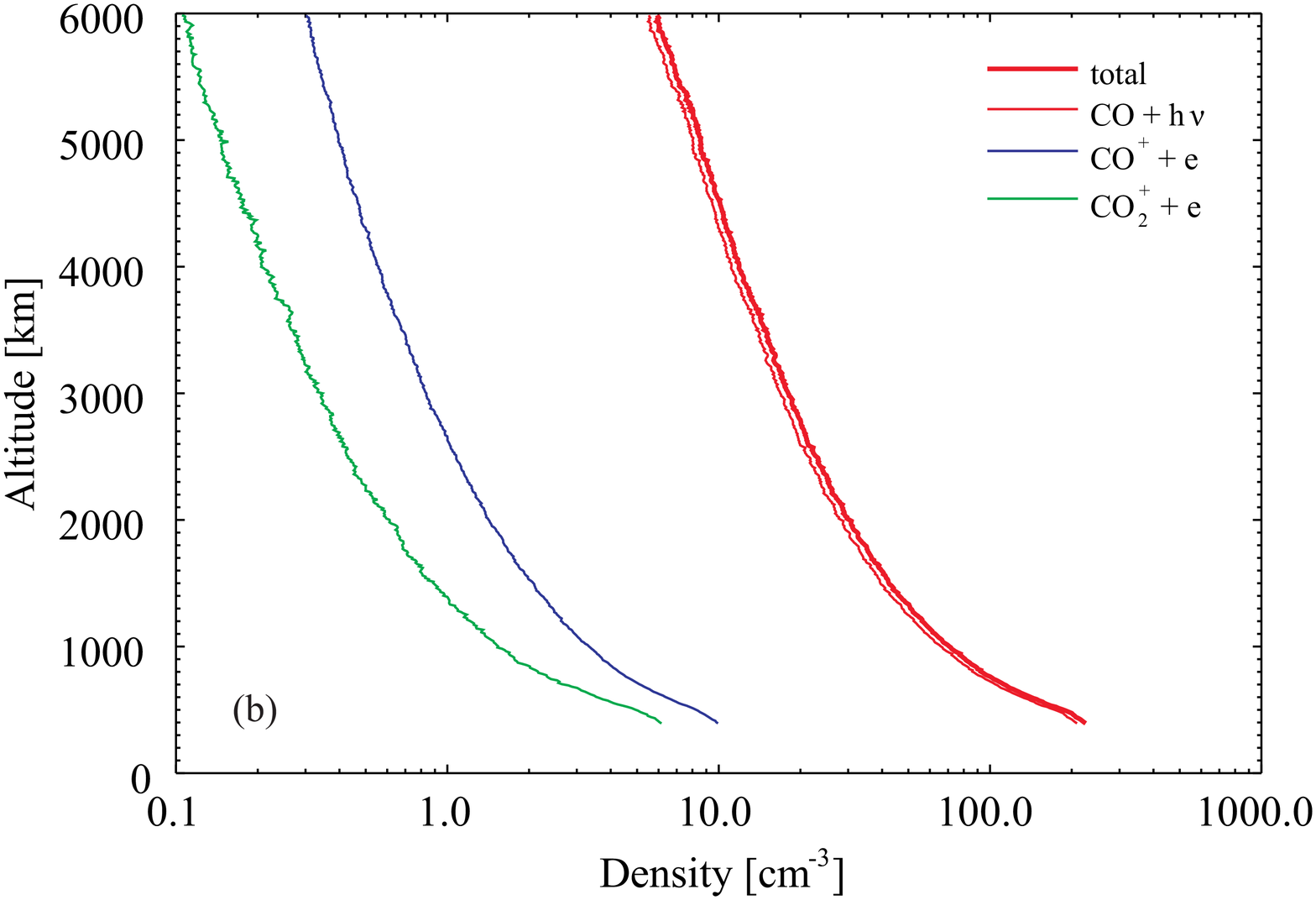}
	\caption{Exosphere densities of hot carbon atoms for various sources for low (a) and high (b) solar activity.}
	\label{fig:density_C}
	\end{center}
\end{figure}

The higher escape rates due to DR of \ce{CO2+} compared to \cite{fox_dissociative_2004} may be somewhat surprising at first glance, since we assume a branching ratio of only 4\% for the \ce{C + O2} channel of DR of \ce{CO2+}, which is lower than the 9\% assumed
by \cite{fox_dissociative_2004}. Moreover, our production rates of carbon are smaller than those calculated in \cite{fox_dissociative_2004} because we use the \ce{CO2+} profiles of \cite{fox_photochemical_2009} which correspond to lower densities than those in \cite{fox_dissociative_2004}. That our escape rates of carbon produced by  DR of \ce{CO2+} are nevertheless higher is at least partially due to the fact that we consider a higher initial velocity for carbon. While \cite{fox_dissociative_2004} assumes that the vibrational-rotational states of the resulting \ce{O2} molecule and the translational states of the C and \ce{O2} fragments are populated statistically, we assume \ce{O2} to be in the electronic and vibrational-rotational ground state, which can be considered as an upper limit for the resulting velocities of the C atoms. In the latter case the velocity distribution peaks at $\sim 1.7$ eV, in the former case at $\sim 0.6$ eV. Another reason for our higher carbon escape rates is due to the different determination of these rates. \cite{fox_dissociative_2004} applies the exobase approximation, where the escape flux is estimated by integrating the production rates above the exobase. Our estimation, however, is based on the integration of the energy distribution function, yielding higher escape rates than the former method -- especially in case of flat production rate profiles like the one for \ce{CO2+ +e}.

The exosphere density profiles for different source terms of the hot C atoms are shown in Figure 7 for low and high solar activity. In Figure 8 the carbon exosphere density obtained by a number of studies is illustrated. It should be noted that \cite{nagy_hot_carbon_2001} include photodissociation of CO, dissociative recombination of \ce{CO+} and collisions between energetic oxygen and thermal carbon atoms as sources of hot C, while \cite{cipriani_martian_2007} consider only dissociative recombination of \ce{CO+}.

\begin{figure}[t]
	\begin{center}
	\includegraphics[width=0.75\columnwidth]{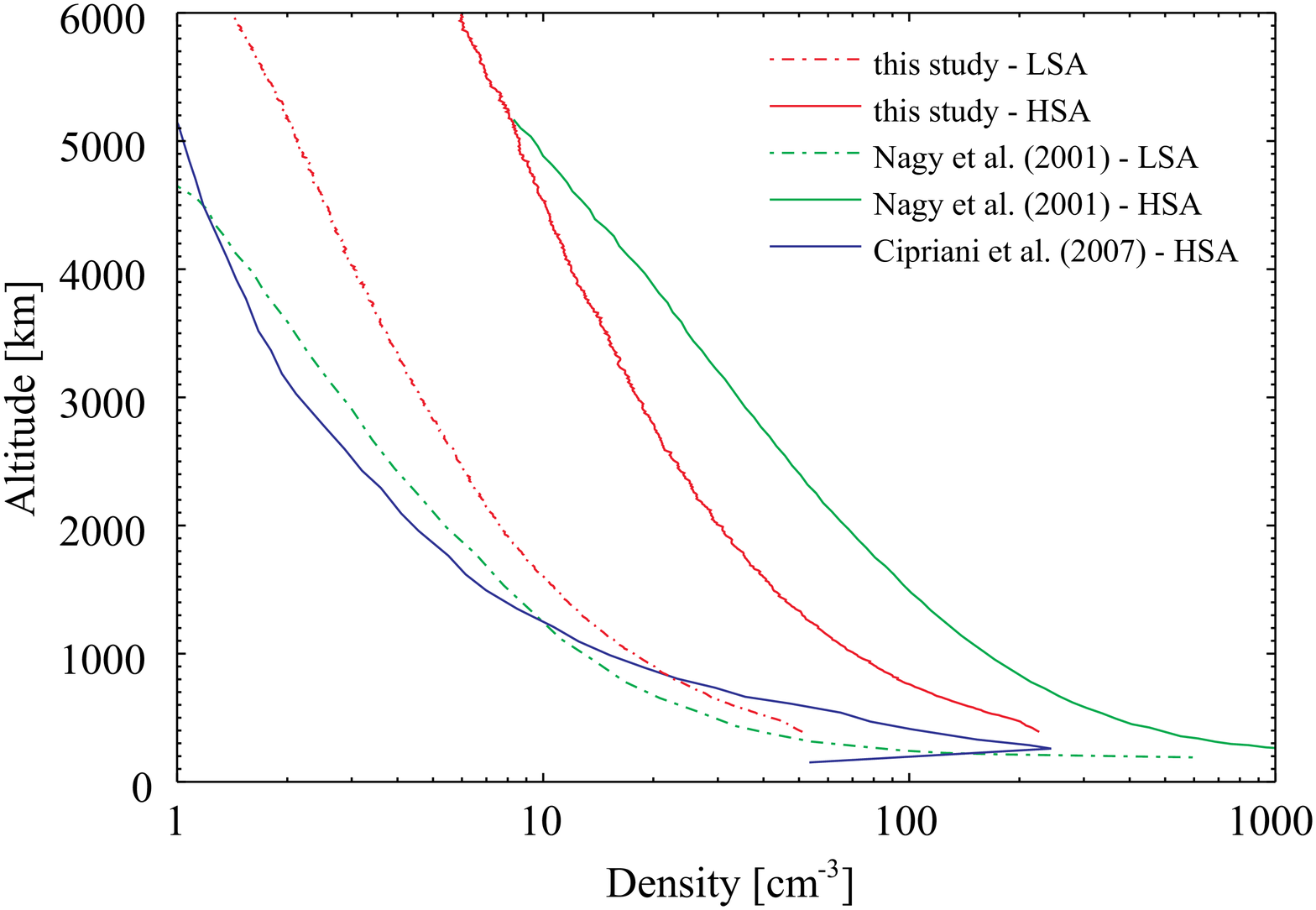}
	\caption{Comparison of the obtained exosphere densities of hot carbon atoms with previous studies.}
	\label{fig:density_C_comparison}
	\end{center}
\end{figure}

Our results suggest -- in agreement with previous studies -- that the main loss of atomic carbon at present Mars is due to photodissociation of CO. According to the present simulations, the total carbon loss is about 0.8 and $3.2\times 10^{24}\,\rm s^{-1}$ for low and high solar activity (Table \ref{tab:C_escape}) , respectively, which is $\sim 5 - 15$ times higher compared to the estimated average atmospheric sputtering of C atoms \citep{chassefiere_combined_2007} and up to $\sim 40$ times higher compared to the estimated \ce{CO2+} molecular ion escape from ASPERA-3 \citep{barabash_martian_2007}.
The present results are in accord with the suggestion of \cite{lammer_outgassing_2013} that the escape of photochemically produced
suprathermal C atoms, which originate from the dissociation of \ce{CO2+}, \ce{CO+} and CO are the most efficient processes for the loss of the martian \ce{CO2} atmosphere at present. Therefore, these exothermic reactions may play a major role for the escape of the martian atmosphere since the end of the Noachian. It is important to note that the modeled hot C and O and the related \ce{CO2} losses depend on a complex interplay of many physical and chemical processes -- e.g. the change of the solar XUV flux and the corresponding response of the upper neutral and ionized atmosphere -- and that the present escape rates can not be easily extrapolated backwards during the planet's history.

Since \ce{CO2+ +e} dissociates mainly into CO and O and only with 4\% probability into C and \ce{O2} (see Table \ref{tab:DR_CO2plus_branches}), it follows O/C=24/1. On the other hand, the corresponding ratio of the escape rates (for e.g. LSA) is $1.0\times 10^{25}/(0.8\times 10^{23})=125$, i.e. oxygen can escape five times more efficient than carbon. This imbalance in the loss rates of O and C can be at least partially attributed to the difference in the energies released by the two channels producing atomic oxygen and carbon (Table \ref{tab:DR_CO2plus_branches}), providing oxygen with a 3.3 times higher energy than carbon.

\section{Summary and Conclusion}
%-------------------------------
A number of possible sources for hot C and O atoms and the resulting escape rates during low and high solar activity conditions from present Mars have been studied by applying a Monte-Carlo model to simulate the stochastic motion of these particles through the upper atmosphere.
The main sources of escaping oxygen are dissociative recombination of \ce{O2+} and \ce{CO2+} leading to a total loss of $\sim 2-3\times 10^{25}\,\rm s^{-1}$ over the solar cycle when an eroded ionosphere model is assumed.
The vast majority of escaping carbon is delivered through photodissociation of CO with a total loss of the order of 0.5 to $3\times 10^{24}\,\rm s^{-1}$ over the solar cycle.
This value is distinctly higher than the estimated present loss due to atmospheric sputtering and the observed \ce{CO2+} escape at Mars.

The calculated loss rates based on the energy distribution function at $450\,{\rm km}$ altitude are reduced by $\sim 20-30$\% with respect to the values calculated by means of the EDFs taken at the exobase level. This suggests that collisions still play a role above the exobase in a region with an extension of $\sim 5-10$ times the local scale height.

It should also be emphasized that the modeled oxygen loss is due to the reactions listed in Tables \ref{tab:DR_O2plus_branches}-\ref{tab:O_sources_reaction} -- above all due to dissociative recombination of \ce{O2+} and \ce{CO2+} -- and is therefore not directly related to the loss of water.
On the other hand, the O escape cannot be balanced by C escape since the former appears to be 10--30 times higher than the latter. This suggests -- assuming the martian atmosphere to be in steady state over long time scales -- that all oxygen lost by escape may be finally replenished by means of \ce{H2O} via a hydrogen-carbon chemistry rather than converting \ce{CO2} into \ce{CO}, thereby stabilizing the \ce{CO2} atmosphere.

\section*{Acknowledgement}
%-------------------------
The authors wish to thank two anonymous referees, whose constructive comments helped to improve the paper.
H. Gr\"oller, H. Lammer and H. Lichtenegger acknowledge support from the Helmholtz Alliance project ''Planetary Evolution and Life'', H. Gr\"oller and H. Lichtenegger also acknowledge support by the Austrian Science Fund (FWF) under the project P24247-N16.
H. Lammer acknowledges also the support by the FWF NFN project S116 ``Pathways to Habitability: From Disks to Active Stars, Planets and Life'', and the related FWF NFN subproject, S116607-N16 "Particle/Radiative Interactions with Upper Atmospheres of Planetary Bodies Under Extreme Stellar Conditions". V. Shematovich acknowledges support from the Basic Research Program of the Presidium of the Russian Academy of Sciences (Program 22), Russian Foundation for Basic Research (Project 11-02-00479a), Federal Targeted Program "Science and Science Education for Innovation in Russia 2009-2013".

%\bibliographystyle{elsarticle-harv}
%\bibliography{references}

%\end{document}

\end{document}